# History-dependent dynamics in a generic model of ion channels – an analytic study

## Daniel Soudry[1,2] and Ron Meir[1,2]*

[1] Department of Electrical Engineering, Technion, Haifa, Israel
[2] Laboratory for Network Biology Research, Technion, Haifa, Israel

Recent experiments have demonstrated that the timescale of adaptation of single neurons and ion channel populations to stimuli slows down as the length of stimulation increases; in fact, no upper bound on temporal timescales seems to exist in such systems. Furthermore, patch clamp experiments on single ion channels have hinted at the existence of large, mostly unobservable, inactivation state spaces within a single ion channel. This raises the question of the relation between this multitude of inactivation states and the observed behavior. In this work we propose a minimal model for ion channel dynamics which does not assume any specific structure of the inactivation state space. The model is simple enough to render an analytical study possible. This leads to a clear and concise explanation of the experimentally observed exponential history-dependent relaxation in sodium channels in a voltage clamp setting, and shows that their recovery rate from slow inactivation must be voltage dependent. Furthermore, we predict that history-dependent relaxation cannot be created by overly sparse spiking activity. While the model was created with ion channel populations in mind, its simplicity and genericalness render it a good starting point for modeling similar effects in other systems, and for scaling up to higher levels such as single neurons which are also known to exhibit multiple time scales.

Keywords: ion channels, slow inactivation, multiple timescales, history-dependence, non-Markov, semi-Markov, renewal theory

## INTRODUCTION

Many recent experiments have demonstrated that the timescale of adaptation of a single neuron in response to periodic stimuli slows down as the period of stimulation increases (Fairhall et al., 2001; Lundstrom et al., 2008; Wark et al., 2009). At a sub-neuronal level, experiments on sodium (Toib et al., 1998; Melamed-Frank and Marom, 1999; Ellerkmann et al., 2001) and calcium (Uebachs et al., 2006) ion channel populations have shown that the timescale of the recovery from inactivation following a long duration of membrane depolarization increased with the length of the depolarization period. We refer to this type of behavior as *history-dependence*. Finally, patch clamp experiments on single ion channels have hinted at the existence of a large inactivation state space within a single ion channel (Liebovitch and Sullivan, 1987; Millhauser et al., 1988; Marom, 1998 and the references therein). These multi-level experimental findings raise several important questions. How are the behaviors observed at the different levels related (e.g., Lowen et al., 1999)? Specifically, is there a connection between the history-dependent timescale of adaptation in the neuron to the history-dependent behavior of ion channels? Does a multitude of inactivation states create the observed channel behavior? What is the functional significance of this history-dependent behavior (e.g., Wark et al., 2009)?

Although we do not address all these questions in this paper, we believe that in order to begin addressing them we first need to construct a simple working, and mathematically tractable, model of slow inactivation in ion channels. Such a model must reproduce the long term behavior in channel population experiments. Our main focus here is the experiment in Toib et al. (1998), which was performed on transfected oocytes and provides very clear empirical findings. In that experiment a membrane with a population of sodium channels of a single type was clamped at low voltage (−90 mV), then at high voltage (−10 mV), for varying lengths of time – from 10 ms to 5 min. During the high voltage stimulus, the sodium channels entered inactivation. Since the fraction of inactivated channels determines the membrane conductivity, by measuring the membrane current, it was possible to observe the dynamics of slow inactivation and recovery in the channel population. After stimulating the membrane with the high voltage clamp for a duration of $t_{stim}$ seconds, the voltage was decreased and clamped back at the low value (−90 mV). At this low voltage level the channels recovered from inactivation. For short stimulations ($t_{stim}$ < 1 s), the recovery was exponential with a single non-history-dependent timescale. After sufficiently long stimulations ($t_{stim}$ > 1 s) the recovery was distinctly exponential and history dependent, the timescale of recovery monotonically increasing with the length of the inactivation period, as seen in Toib et al. (1998). Interestingly, early experiments on visual receptors already displayed similar behavior (Baylor and Hodgkin 1974; in particular see Figures 18A,B).

This history-dependence is generally thought to result from the large inactivation state space hinted at by the single channel patch clamp experiments, as suggested first by Toib et al. (1998). Previous modeling approaches, based on this idea, have already been suggested in the literature, but fall short in accurately reproducing this behavior. We present a comparative discussion in Section 'Relation to Previous Work'. One difficulty in modeling channel behavior is that the nature of the protein conformation dynamics leading to the complex properties of ion channels at long timescales is currently ill-understood, precluding the construction of a full bottom-up biophysical model of ion channels. In fact, it is unclear whether such dauntingly complex low-level models would be useful in explaining phenomena at the level of current interest.

In this work we present a simple two-state generic mathematical model which requires very few assumptions on the nature of the inactivation state space, and which leads to concise explanations of observed experimental findings, and to concrete predictions for future experiments. We reproduce for the first time, to our knowledge, the main experimental finding from Toib et al. (1998), namely an exponential recovery process with a history-dependent timescale, as demonstrated in **Figure 3**. Using these results and other similar experiments (Ellermann et al., 2001; Hering et al., 2004; Uebachs et al., 2006), we narrow down the options for the model parameters at different voltages for several channel types, and explicitly address the issue of long-memory phenomena (Mercik and Weron, 2001). The model introduced here also provides many predictions. Qualitatively, we predict that temporally spaced spiking stimuli will have a significantly reduced effect on the timescale of channel recovery from inactivation, and that the rate of this recovery must be voltage dependent. Quantitatively, we derive a dynamic equation that fully defines an input–output relation between the membrane voltage and channel availability and solve it exactly in many important cases. Additionally, we develop expressions that describe all joint moments in the single channel and population.

We note that the potential contribution of this model goes beyond the specific system addressed in this work. As pointed out in Marom (2009, 2010), current models of channels and receptors (e.g., Faber et al., 2007) tend to suffer from an embarrassment of riches. In order to explain behaviors over an ever expanding range of timescales, these complex models often include multiple inactivation states. Since the number of states and their parameters are not directly observable, these models tend to be highly specific and are likely to suffer from over-fitting. Furthermore, such models always have an upper bound on their timescale. In this work, we introduce and thoroughly analyze, for the first time to our knowledge, a type of model that does not suffer from these limitations. Despite its simplicity, it provides a generalization of previous models, is based only on measurable quantities, does not possess an upper bound on its timescale and exhibits considerable analytical tractability. As such, it stands as an appealing alternative to previous approaches, and as a basic building block in the construction of higher neuronal models. For example, the work of Lowen et al. (1999) clearly demonstrates the strong impact of a similar model at the channel level on the long term statistics of neuronal firing.

The outline of the article is as follows. We begin in Section 'The Model' by motivating the model and describing its structure, and then present several exact solutions for the channel dynamics for different types of voltage inputs in Section 'Response of Channel to Different Types of Voltage Input'. In Section 'Reproduction of Experimental Results' we demonstrate how to reproduce experimental results from Toib et al. (1998), then we use these results to narrow down the possible parameter values, and make specific predictions. In Section 'Temporal Correlations' we discuss temporal correlations and long-memory, and in Section 'Relation to Previous Work' we explain the relation of this paper to previous theoretical work. Next, we summarize and discuss our results in Section 'Discussion', and present the mathematical details of the analytical methods used in Section 'Methods'. In the Supplementary Material we address several technical issues and provide the simulation code.

## RESULTS
### THE MODEL
#### Background

The total conductivity, $G(t)$, of a population of ion channels located on a piece of membrane is determined by $A(t)$, the portion of channels which are available for conducting, through the equation $G(t) = G_{max}A(t)$, where $G_{max}$ is the conductivity of the membrane when all channels are available. In the limit of many channels which are independent *given* the voltage, the law of large numbers implies that $A(t)$ is approximately equal to $p(t)$, the probability of a single channel to be available (see Supplementary Material). Thus, in order to understand the conductivity dynamics of a population of channels, we first calculate the probability of a single channel to be available, given the input voltage.

Ion channels are commonly modeled as a continuous-time Markov process with a finite, possibly large, number of states (Colquhoun and Hawkes, 1977). Each state is completely characterized by the probability density function of its residence time (the time required to leave that state) and the probabilities to shift to other states when this transition occurs. This division into 'states' implies that the dynamics of a single channel following a transition from one state to another is independent of the history prior to that transition. The term 'Markov process' implies that this 'memoryless' behavior is also maintained in each state, namely, the transition out of the state is independent of the time the channel already resided in that state. Slightly abusing notation, we refer to a state as 'Markovian' if the dynamics in that state are memoryless. For a Markovian state, the transition rates leading out of this state are constant, and the resulting residence time probability density function (RTPDF) is exponential.

In these common Markovian models, the states are divided into groups of 'open', 'closed' and 'inactivated' states. The channel may conduct ions only in an 'open' state, thereby actively participating in the generation of an action potential. As noted in Marom (1998) and Marom (2009), it is possible to lump together into a single Markovian 'available' state all the states from which the channel may change into 'open' quickly enough to participate in the creation of a single action potential. This is possible because transitions between the states that compose the available state are much faster than the transitions to and from non-available states (see Supplementary Material). In contrast, if we lump together all the remaining slow inactivation states into a single state, generally it is not a Markovian state. Therefore, we need to switch from the familiar concept of a Markov process to the more general 'semi-Markov process' in which some of the states may be non-Markovian (for exact definitions and properties, see Cinlar, 1975).

#### Model description

In view of the above comments, motivated by Liebovitch (1989) and Marom (2009), we model the slow inactivation of a channel as a continuous-time semi-Markov process consisting of only two states, the Markovian 'available' state, and the non-Markovian 'inactivated' state, as shown in **Figure 1**.

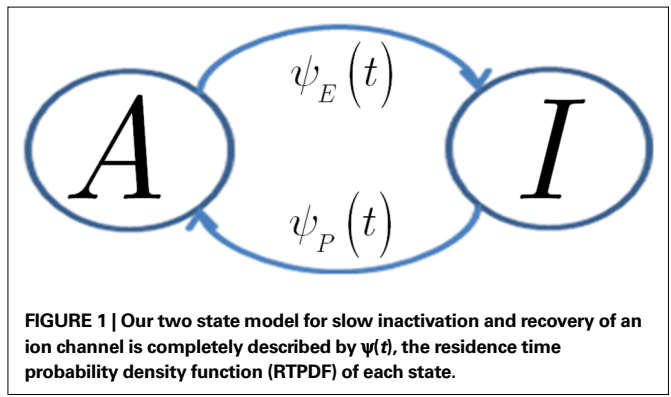

**FIGURE 1 | Our two state model for slow inactivation and recovery of an ion channel is completely described by ψ(t), the residence time probability density function (RTPDF) of each state.**

The RTPDF of the Markovian available state is exponential with parameter $\gamma$,

$$\psi_E(t) = \gamma \exp(-\gamma t), \quad t \geq 0. \tag{1}$$

The inactivated state is non-Markovian, where we use the power-law RTPDF:

$$\psi_P(t) = \frac{c/t_0}{(1 + t/t_0)^{c+1}}, \quad t \geq 0, \tag{2}$$

where $\gamma$ and $c$ are, in general, voltage-dependent, while $t_0$ is taken to be constant. The way the RTPDFs change with voltage is explained in Section 'General Voltage Input'. From the normalization demand imposed on the RTPDF, $\gamma$, $c$ and $t_0$ must all be strictly positive constants.

We comment that although the structure of the model is indeed very simple, it gives rise to an unexpected richness of behaviors as a function of the underlying parameters. This exact and detailed mathematical analysis, covering a broad range of parameter values, is presented here for the first time. Moreover, subtle mathematical issues arise in correctly characterizing the different regimes. None of these complexities appear when dealing with the more standard Markovian channel models.

### Model motivation

Here we choose to model the, possibly complex, slow inactivation state space solely through a single non-Markovian state and its associated RTPDF. As stated above, the transition structure and rates by which the channel proteins change conformations between different slow inactivation states are still ill-understood. Moreover, the only quantity related to this state space that may be directly measured is the RTPDF. Previous single channel patch clamp experiments have already measured the RTPDFs of different channels. In many cases single channel patch clamp experiments have indeed fitted their measured RTPDF for the closed time with a similar power-law function as above. However, those experiments were performed on timescales of milliseconds-to-seconds. Thus, at this experimental stage, we can only speculate as to whether the RTPDF is still a power law on the timescale of seconds-to-minutes, which is the relevant timescale in Toib et al. (1998). Our reasons for using the particular two-parameter power-law RTPDF in Eq. 2 are the following.

Any smooth normalizable RTPDF $\psi(t)$ must decay to zero as $t \to \infty$. The asymptotic form of $\psi(t)$ as $t \to \infty$ is referred to as the 'tail of the distribution'. In experiments done on channels at long timescales, such as Toib et al. (1998), this tail dominates the dynamics of the channel. We sought to investigate a simple RTPDF with a non-exponential tail behavior, allowing for a non-Markovian inactivated state, which may lead to a history-dependent behavior. The power-law RTPDF function in Eq. 2 is the simplest possible form. Such power-law RTPDFs frequently appear in models of disordered systems, e.g., spin-glasses (Bouchaud, 1992). Moreover, a similar power-law RTPDF was used in Lowen et al. (1999) to successfully simulate long term temporal behaviors at the level of a single neuron (in that model a specific voltage-dependent choice of $c$ was made, and the available state had also a power-law RTPDF, rather than exponential).

The parametrization Eq. 2 includes two important elements – $c$ as a parameter that determines the tail of the distribution, and $t_0$ as a lower boundary on the temporal resolution. For simplicity, we chose it to be constant. It is interesting to note that in limit $c \to \infty$ $\psi_P(t)$ approximates an exponential RTPDF corresponding to a Markovian state (see Supplementary Material).

### RESPONSE OF CHANNEL TO DIFFERENT TYPES OF VOLTAGE INPUT

In this part we present several results on the behavior of the channel in response to different types of voltage input. All of the these results are derived and proved using analytical techniques (see Methods) and demonstrated numerically.

### Relaxation to a steady state under a step voltage

First we investigate what happens to the channel when the voltage is maintained fixed, so that $c$ and $\gamma$ are constant. By 'projecting' the single inactivated state onto a continuum of Markovian states, we derive in Eq. 16 a dynamic equation for $p(t)$, the probability of occupying the available state,

$$\frac{d}{dt} p(t) = -\gamma p(t) + \int_0^t \gamma p(t') \psi_P(t - t') dt'. \tag{3}$$

Here we used the initial condition $p(0) = 1$; an extension to general initial conditions is presented in Eq. 16. The intuition behind this equation is simple: the first term represents the loss of probability from the available state, while the second term represents the probability current that goes from availability to inactivation and back again, where we integrate over all the possible past inactivation times. Note that $\psi_P(\cdot)$ can generally be replaced in this equation by any RTPDF of the inactivated state, but we focus in our analysis on the case of the power-law RTPDF from Eq. 2.

We solve Eq. 3, and find that $p(t)$ relaxes asymptotically in a power-law manner, to a steady state value,

$$p(t) = p_\infty + (1 - p_\infty) q t^{-|1-c|}, \tag{4}$$

where $q = \sin(\pi c)/(\pi \gamma t_0^c)$ for $0 < c < 1$, $q = p_\infty t_0^{c-1}$ for $c > 1$, and $p_\infty$ is the steady state value:

$$p_\infty = \begin{cases} \frac{c-1}{\gamma_0 + c - 1} & \text{if } c \geq 1, \\ 0 & \text{if } 0 < c < 1. \end{cases} \tag{5}$$

We can express $p_\infty$ in the more familiar form of the steady state of a simple two-state Markov process, $p_\infty = \tilde{\delta}/(\tilde{\delta} + \gamma)$ where $\tilde{\delta}$ is, as in the Markov scheme, the inverse of the mean inactivation time:

$$\tilde{\delta} = \left( \int_0^\infty t\psi_P(t)dt \right)^{-1} = \begin{cases} 0 & \text{if } c \leq 1, \\ \frac{c-1}{t_0} & \text{if } c > 1. \end{cases} \tag{6}$$

In **Figure 2** we compare these asymptotic results with a numerical simulation. The agreement between the two results in this case indicates that the asymptotic results are accurate for a wide range of timescales.

Notice that for $c \leq 1$, for any choice of the other parameters, $p(t) \to 0$. In other words, the channel eventually decays to complete inactivation, independently of the value of the other parameters (recall that $\gamma > 0$). Intuitively, in this case the residence time mean is infinite, and the rate of return to the available state is so slow, that ultimately it remains unoccupied. Assuming that real channels at rest voltage do not decay to complete inactivation, we conclude that $c > 1$ at rest voltage, namely $c(V_{rest}) > 1$.

### History-dependent recovery timescale in response to voltage pulse

In Section 'Relaxation to a steady state under a step voltage' we studied the channel dynamics during a constant voltage step. The major feature of the model introduced here is its history-dependent dynamics, a notion which cannot be investigated for such a simple input. In order to quantify this notion, in the present section we consider a voltage pulse of constant amplitude and finite duration $t$, studying the recovery process immediately following the termination of the pulse (similarly to was done experimentally in Toib et al., 1998).

Consider a single channel at time $t$ (the end of the voltage pulse). When the voltage is changed abruptly from one fixed value to another, $c$ and $\gamma$ also change. If during this voltage change the channel is in the available (Markovian) state, it does not 'remember' its history prior to the voltage change. If the channel is in the inactivated state during the voltage change, the subsequent dynamics depends on the prior history of the channel through the variable $T$ – the duration of the channel's sojourn in the inactivated state at time $t$. For the inactivated channel, the probability to recover at times between $t$ and $t + dt$ depends on $T$. This probability, divided by $dt$, is termed the *time-dependent rate of recovery*, and its inverse is the *time-dependent timescale of recovery*, $\tau$. A simple derivation (Eq. 35) shows that $\tau$ depends linearly on $T$,

$$\tau = \frac{T + t_0}{c}. \tag{7}$$

And so, immediately after a voltage change, the timescale of recovery is determined by Eq. 7. In this model the variable $c$ changes instantly with voltage, while $T$ is a continuous variable which retains the same value it had prior to the change. We note here that it is possible to model the effect of the voltage change on the time-dependent rate of recovery differently (see General Voltage Input).

For each channel, $T$ is a random variable, and so in a population of channels, $T$ has some distribution of values. We develop in Eq. 36 an exact asymptotic expression for this distribution, in the voltage pulse setting, where we set the initial condition so that all channels are initially available. We define $\langle T \rangle$, the mean duration of a channel in the inactivated state, and $CV_T \triangleq \sigma_T / \langle T \rangle$, the coefficient of variation of the distribution of $T$, given by the ratio between the standard deviation and the mean of the distribution of $T$. The latter variable measures the dispersion of the distribution of $T$ around its mean.

To approximate the timescale of recovery of the channel population after the voltage pulse, we can substitute $T$ by $\langle T \rangle$ in Eq. 7. If $CV_T \ll 1$ then this approximation is accurate, and the recovery after the pulse will follow an exponential form, to a good approximation. If $CV_T \gg 1$ this approximation will not be accurate, since then the mean provides a poor representation of the distribution. Also, when $CV_T \gg 1$ the recovery will be distinctly non-exponential.

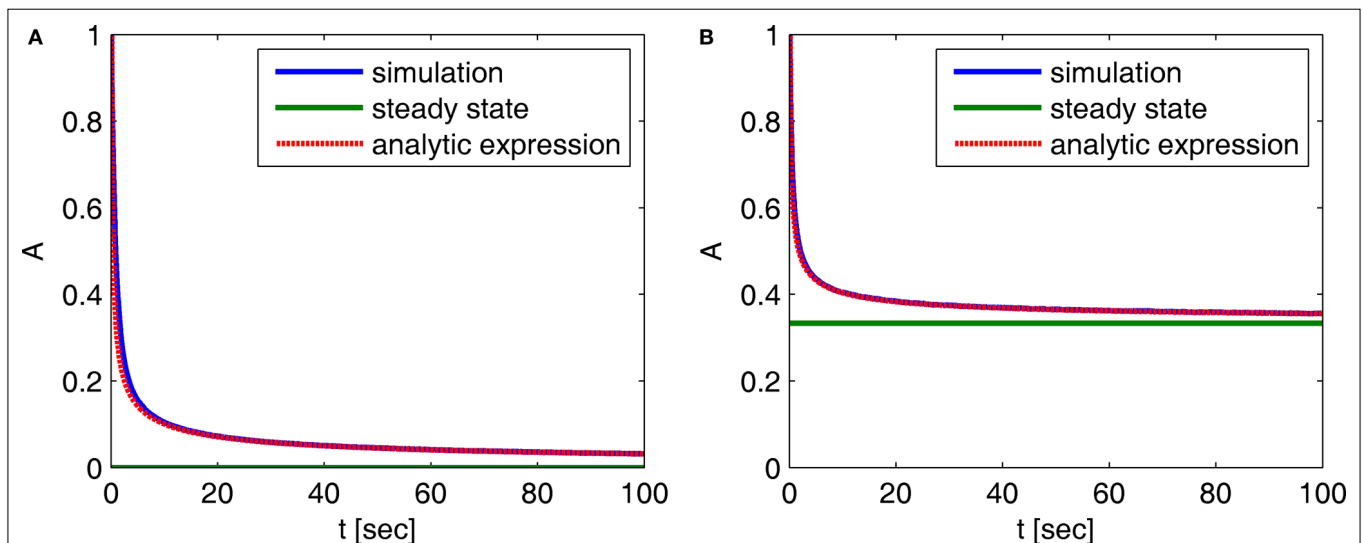

**FIGURE 2 | Behavior of $A(t)$ – comparison between numerical simulation and the analytical expression of the asymptotic behavior of $A(t)$: (A) $c = 0.5$, (B) $c = 1.5$.** The remaining parameters are: $N = 10^6$, $\gamma = 1$ Hz, $t_0 = 1$ s and a simulation step of $\Delta t = 5 \times 10^{-3}$ s. The agreement between the analytical and numerical results in this case demonstrates that the asymptotic analytical results are accurate for a wide range of timescales.

In Section 'Renewal Theory Approach' we compute exactly the behavior of $\langle T \rangle$ and $CV_T$ at the end of the voltage pulse, at time $t$. This behavior depends in an intricate way on the value of the parameter $c$ during the pulse. And as we explained, the value of $\langle T \rangle$ and $CV_T$ at the end of the pulse determines, through Eq. 7, the manner in which the channels recover immediately after the pulse:

- $c < 1$: $\langle T \rangle = (1 - c)t$, $CV_T = \sqrt{c/(2(1 - c))}$. The recovery time-scale increase linearly with $t$, while maintaining a constant dispersion. Also, Since $CV_T$ increases monotonically with $c$, the distribution of timescales becomes less dispersed for lower values of $c$, and hence, the recovery following the pulse becomes more exponential.

- $1 < c < 3$: $CV_T \to \infty$. Thus, the distribution of recovery timescales broadens constantly, which entails a non-exponential recovery (in this case, the mean, $\langle T \rangle$, provides a poor representation of the distribution).

- $c > 3$: $\langle T \rangle = t_0/(c - 2)$, $CV_T = \sqrt{(c - 1)/(c - 3)}$. Since both $\langle T \rangle$ and $CV_T$ converge to a fixed finite value, we get that the recovery following the pulse takes place at a single timescale. Higher values of $c$, lead to more exponential recovery. This implies that in this case the inactivated state is 'almost Markovian'.

The (technical) case of integer valued $c$ is discussed in the Supplementary Material. Observe that the qualitative change of behavior noted above, results from the order of the first infinite moment of the RTPDF $\psi_p$, which depends on the value of $c$. More specifically, the first moment diverges for $c \leq 1$, the second moment diverges for $c \leq 2$, and so on.

### Complex voltage input

So far, we have described the asymptotic channel dynamics when the voltage is constant, and also immediately after the voltage has jumped from one constant value to another. Here we briefly discuss two other cases in which the channel dynamics may be solved analytically, and then discuss the general case of time-varying voltage.

The first case corresponds to the adiabatic limit, when the voltage changes on a timescale which is far slower than the timescales of the channel, namely the timescale of inactivation $\gamma^{-1}$, and the timescale of recovery $\tau$. Notice that the timescales are themselves voltage dependent, and $\tau$ may even be history dependent, so it is not always trivial to determine whether we are indeed in the adiabatic limit. In this limit, we may assume that both $p(t)$ and the distribution of $T$ follow their steady states.

The second case occurs in the opposite limit, in which the voltage oscillates rapidly around some constant value with an effective period of $T_p$. By using the word 'effective' here we allow the oscillation to be stochastic (noise) with a period which is only approximate. In this case, we show in Eqs 55 and 56 that we can replace $\gamma$ and $c$ with the 'effective parameters' $\hat{\gamma}$ and $\hat{c}$ given by the time-averaged values of $\gamma$ and $c$, respectively,

$$\dot{\hat{\gamma}} = \frac{1}{T_p} \int_0^{T_p} \gamma(V(t)) \, dt \; ; \quad \hat{c} = \frac{1}{T_p} \int_0^{T_p} c(V(t)) \, dt, \quad (8)$$

where $V_t$ is the voltage at time $t$. Then we may again use the results derived in the case of constant voltage.

Finally, when none of these approximations is valid, we show in Eq. 43 that it is possible to derive a closed form integro-differential equation generalizing Eq. 3 to the case of arbitrary input.

## REPRODUCTION OF EXPERIMENTAL RESULTS

As observed in Section 'Response of Channel to Different Types of Voltage Input', the channel dynamics depends intricately on the parameter $c$. In the present section we consider experimental results relating to channel dynamics, in order to test the predictions of our model, and, importantly, constrain the possible values of $c$.

### Exponential and history-dependent relaxation in response to long depolarizations

Using the model presented, we wish to reproduce the main experimental results in Toib et al. (1998). In this experiment, the membrane is clamped at a high voltage of −10 mV for varying length of time, $t_{stim}$, and then clamped at low voltage of −90 mV – in which the recovery from inactivation occurred. This recovery was exponential and history dependent for every stimulus longer than 1 s. Analyzing these experimental observations using our model, we are able to restrict the possible values of the parameter $c$. We denote by $c_H$, $\gamma_H$ and $c_L$, $\gamma_L$ the model's parameter values during the high and low voltage phases, respectively. Based on the experimental results from Toib et al. (1998) we make the following claim.

Claim: During inactivation $c = c_H < 1$, while during recovery $c = c_L > 3$.

This claim is based on the characterization of the different regimes of $c$, provided in Section 'History-dependent recovery timescale in response to voltage pulse'. First, we infer that $c_L > 1$. This follows from Eqs 4 and 5, since for $c_L < 1$, the availability of the membrane declines to zero, a phenomenon which did not occur in the experiment. Second, we argue that $c_L$ is not in the range $1 < c_L < 3$, since if it were, the distribution of recovery timescales in the population would keep broadening with time, thus causing the recovery of $A(t)$ to be non-exponential. Thus, we conclude that $c_L > 3$.

Next, we argue that $c_H < 1$. First, it is clear that $c_H < 3$, since from Eq. 7 and the behavior of $\langle T \rangle$ in this case, we know that during the inactivation period, $\langle T \rangle$ is bounded by its steady state value $\langle T \rangle = t_0/(c_H - 2)$. By substituting this value into Eq. 7, and comparing with the result when $\langle T \rangle = 0$, we learn that in this case we should not expect to see large changes in the recovery timescales following different stimuli durations, contradicting the experimental results. Finally, $c_H$ cannot be in the range $1 < c_H < 3$ since, if it were, the distribution in the inactive state following the high voltage period would be very broad (a high value of $CV_T$), rendering an exponential relaxation impossible.

We thus conclude that in order to fit the experimental results, we must set $c_H < 1$ during the inactivation period and $c_L > 3$ during the recovery phase. In this case, the timescale of recovery increases linearly with the stimulus duration, and the recovery is exponential as long as:

1. $c_H$ is small enough so that $CV_T \ll 1$, since $CV_T = \sqrt{c_H / (2(1 - c_H))}$, which is increasing in $c_H$.
2. $c_L$ is large enough in the low voltage phase so that $\langle T \rangle$ does not change considerably during the recovery. More formally, $T \gg \tau = (T + t_0)/c_L$, or more simply, $c_L$ is sufficiently larger than 1.
3. $\gamma_I$ is small enough during the recovery period, so that during this period, the timescale of recovery does not change by 'freshly inactivated' channels; more formally, $\gamma_I^{-1} \gg \tau = (T + t_0)/c_L$.

In this method of reproducing the experimental results the channel possesses 'infinite memory', since as long as the channel remains at the high voltage, $\langle \tau \rangle$ continues to increase linearly:

$$\langle \tau \rangle = \frac{(1 - c_H) \cdot t_{\text{stim}} + t_0}{c_L} \tag{9}$$

We note here that it is possible to modify the model so that $t_0$ is the voltage-dependent parameter and $c$ is fixed at some value ($>1$). In this case, the channel does not possess this 'infinite memory' and the timescale of recovery is bounded. Therefore, in order to reproduce the experimental results in this modified model, we have to assume that the increase in the timescale of recovery has some upper boundary, which was not yet reached in the experiment. Since it seems unnatural to assume the existence of such upper limit, this alternative model was not used.

In an additional experiment the recovery was examined under several voltages: $-60, -90, -120$ mV. The history-dependent behavior remained, as can be seen in Figure 6 in Toib et al. (1998). The recovery was similar for $-90, -120$ mV, indicating that $c$ did not change much between these voltages. At $-60$ mV, the recovery was slower, and possibly less exponential, indicating perhaps that $1 < c_L < 3$, in that case.

A further experimental observation relates to the emergence of a history-dependent relaxation only for $t_{\text{stim}} \geq 1$ s. From this threshold point between the two modes of recovery and the results in Section 'History-dependent recovery timescale in response to voltage pulse' we get that at this experiment $t_0 \sim 1 \, [\text{s}]$.

In any case, the resulting prediction of this model is that the rate of recovery from the inactivated state must decrease with voltage difference between the two values examined in the experiment. This is in accordance with the results of Figure 7 in Fleidervish et al. (1996) and Figure 4 in Ellermann et al. (2001), where it is observed that the timescale of recovery of slow inactivation in sodium channels increases monotonically with voltage. Moreover, if we assume, in accordance with these results, that $c(V)$ is continuous for those sodium channels, a further prediction of our model is that there exists some range of voltages for which $1 < c_H < 3$, where the recovery becomes non-exponential.

### Channel response to a spiking stimulus

Motivated by the results in Ellermann et al. (2001), Toib et al. (1998) and Uebachs et al. (2006), we study the model's dynamics when the voltage input is a periodic voltage spike train. As noted in Toib et al. (1998), such inputs are similar to firing patterns in neocortical neurons. This type of input may be important when considering the effects of a neuron's action potential on itself. We note that such an input does not present the entire picture, since synaptic inputs from other neurons are probably more realistically described as sums of continuous functions (Gerstner and Kistler, 2002). In any event, it is interesting to test the model's prediction in this setting, for which some experimental results are available (Toib et al., 1998; Ellermann et al., 2001; Uebachs et al., 2006). The voltage spikes are modeled here as a square wave – for $T_H$ seconds the voltage is set high, and for $T_L$ seconds

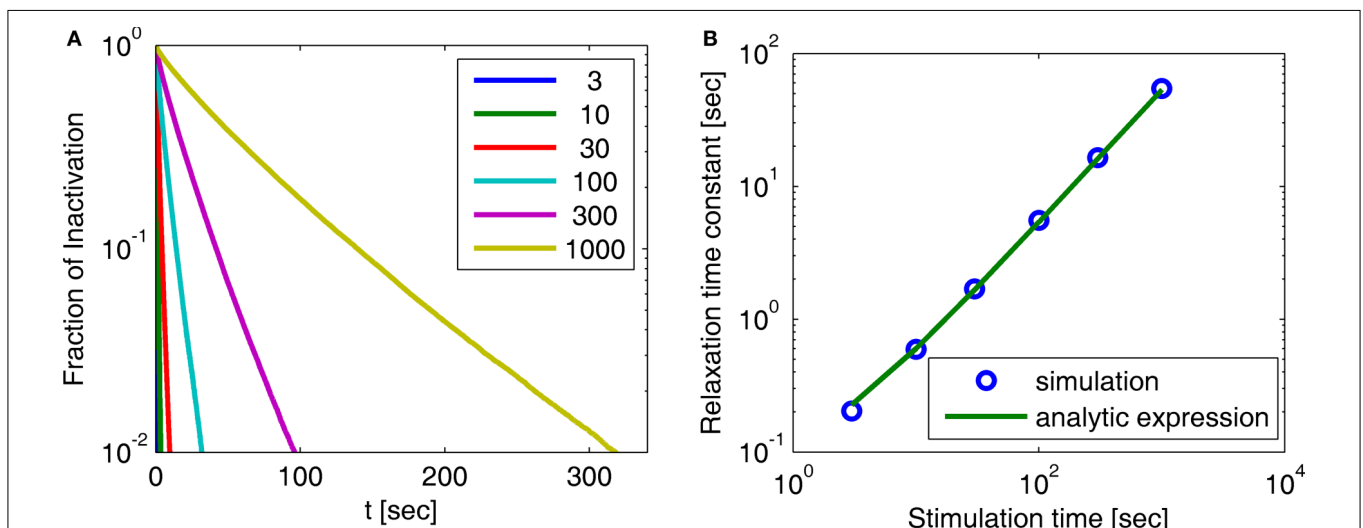

**FIGURE 3 | Simulation of experimental results. (A)** Exponential and history-dependent recovery in response to various stimulus lengths ($R^2 > 0.99$ for all exponential fits). Fraction of inactivation is defined as in Toib et al. (1998). Legend: different stimuli lengths, in seconds. **(B)** The increase in the timescale of recovery with stimulation length, in both the numerical simulation and the analytic expression (Eq. 9) (with $c_H < 1$). We see that the two results agree,

especially for long stimulations. This demonstrates how, in this case, the asymptotic analytic result is a good approximation for the relevant timescales. The linear relationship is expected, based on the analytic result, to persist for arbitrarily long stimuli. The parameters for the simulation were $N = 10^5$, $\gamma_L = 10^{-4}$ Hz, $\gamma_H = 1$ Hz, $c_H = 0.2$, $c_L = 15$, $t_0 = 1$ s and a simulation step of $\Delta t = 10^{-3}$ s.

the voltage is set low. As before, we denote by $c_H$, $\gamma_H$ and $c_L$, $\gamma_L$ the model parameters during high and low voltage, respectively (see **Figure 4**).

Assuming that the input spike frequency is much higher than the transition timescales of the model, we use the effective parameter approximation. The behavior of the system may be inferred by replacing $\gamma$ and $c$ with the effective parameters, according to Eq. 8:

$$\hat{\gamma} = \frac{T_H\gamma_H + T_L\gamma_L}{T_H + T_L}; \quad \hat{c} = \frac{T_H c_H + T_L c_L}{T_H + T_L}. \tag{10}$$

It is important to notice in this case that if $T_H \ll T_L$ (sparse spikes), then $\hat{\gamma}$ will be affected by $\gamma_H$ only if $\gamma_H \gg \gamma_L$, and similarly for $\hat{c}$. Since we know from the experimental results that $c_H < c_L$, in this case we can approximate, $\hat{c} \approx c_L$. Using this approximation we address the dependence of $\gamma$ on the voltage. From Toib et al. (1998) it is known that the sodium channel population goes into significant inactivation as a result of an action-potential-like stimulus. This is also the case in Ellerkmann et al. (2001). Since in this case $\hat{c} \approx c_L$ (does not change much), then $\hat{\gamma}$ must be significantly larger than $\gamma_L$ – so $\gamma_H$ must be significantly higher than $\gamma_L$ (otherwise, the experimentally observed inactivation would not be reproduced). This means that $\gamma(V)$ must increase with voltage, at least between the two voltage values used in this setup.

The distribution of recovery timescales in the channel population is affected mainly by the value of $c$ (the dependence on $\gamma$ is absent from the asymptotic form Eq. 36). If we wish to make the timescale of recovery change measurably in response to spike stimulation, $\hat{c}$ must be significantly different from $c_L$. For example, if a neuron is affected solely by its own action potentials, then $T_H \approx 1$ ms, $T_H + T_L > 10$ ms, and therefore $c_L > \hat{c} > 0.9c_L$. So, in this case it would be hard to experimentally detect an increase of the recovery timescales due to the input (see **Figure 5**). This assertion will remain valid for any type of stimulus in which $T_H \ll T_L$ – the channel may enter inactivation, but any history-dependent recovery will be mitigated. This complies with the results shown in Figure 5 of Ellerkmann et al. (2001), where we can see a relatively weak scaling of the recovery timescale with stimulus frequency – even though $T_H/(T_H + T_L)$ is rather high in comparison with typical action potentials.

It is important to note that in the case $\hat{c} > 3$, the non-Markovian state can be approximated by a simple Markovian state, since in this case $\langle T \rangle < t_0$, and thus $\tau = (T + t_0)/\hat{c}$ does not change

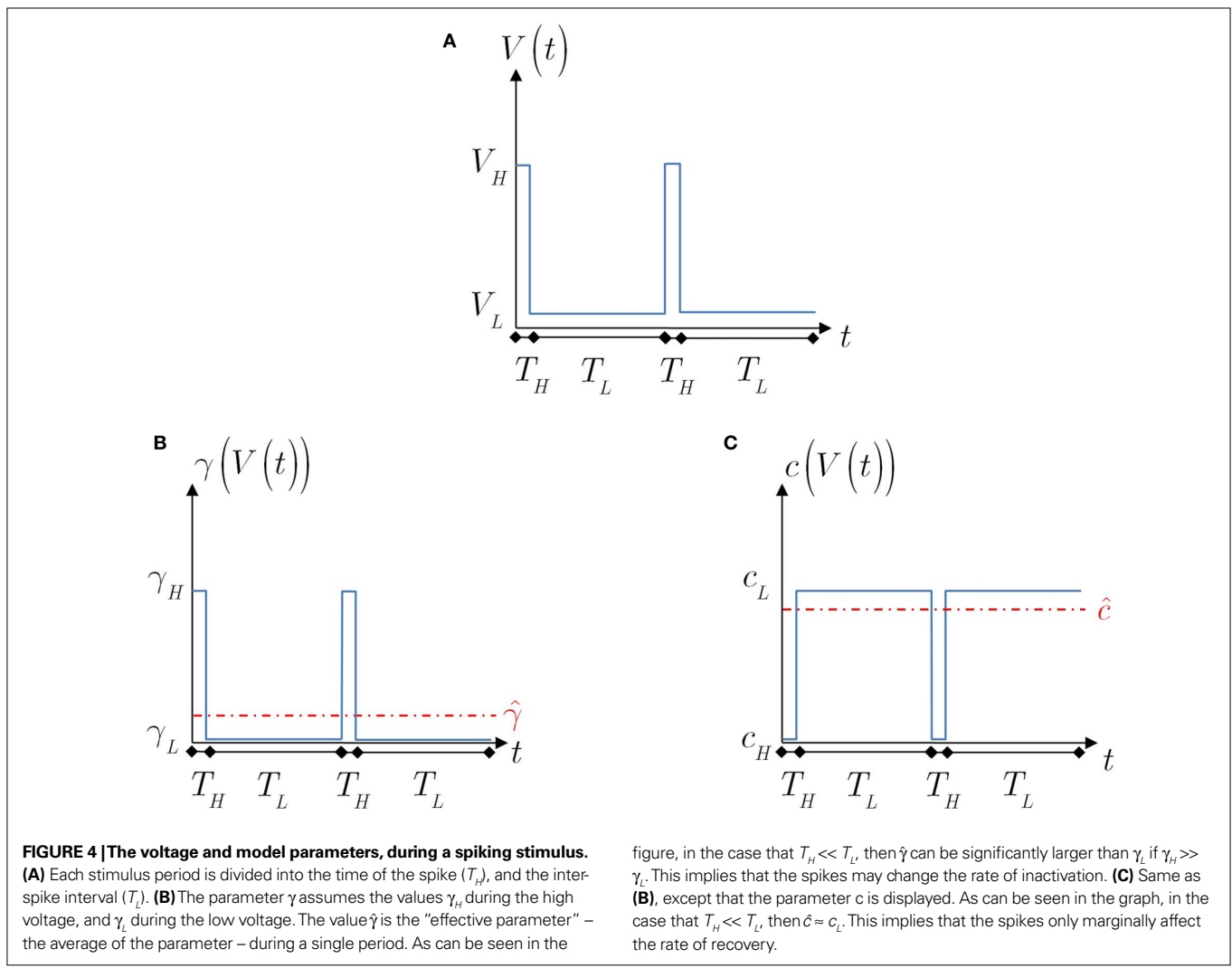

**FIGURE 4 | The voltage and model parameters, during a spiking stimulus.**
**(A)** Each stimulus period is divided into the time of the spike ($T_H$), and the inter-spike interval ($T_L$). **(B)** The parameter $\gamma$ assumes the values $\gamma_H$ during the high voltage, and $\gamma_L$ during the low voltage. The value $\hat{\gamma}$ is the "effective parameter" – the average of the parameter – during a single period. As can be seen in the figure, in the case that $T_H \ll T_L$, then $\hat{\gamma}$ can be significantly larger than $\gamma_L$ if $\gamma_H \gg \gamma_L$. This implies that the spikes may change the rate of inactivation. **(C)** Same as **(B)**, except that the parameter $c$ is displayed. As can be seen in the graph, in the case that $T_H \ll T_L$, then $\hat{c} \approx c_L$. This implies that the spikes only marginally affect the rate of recovery.

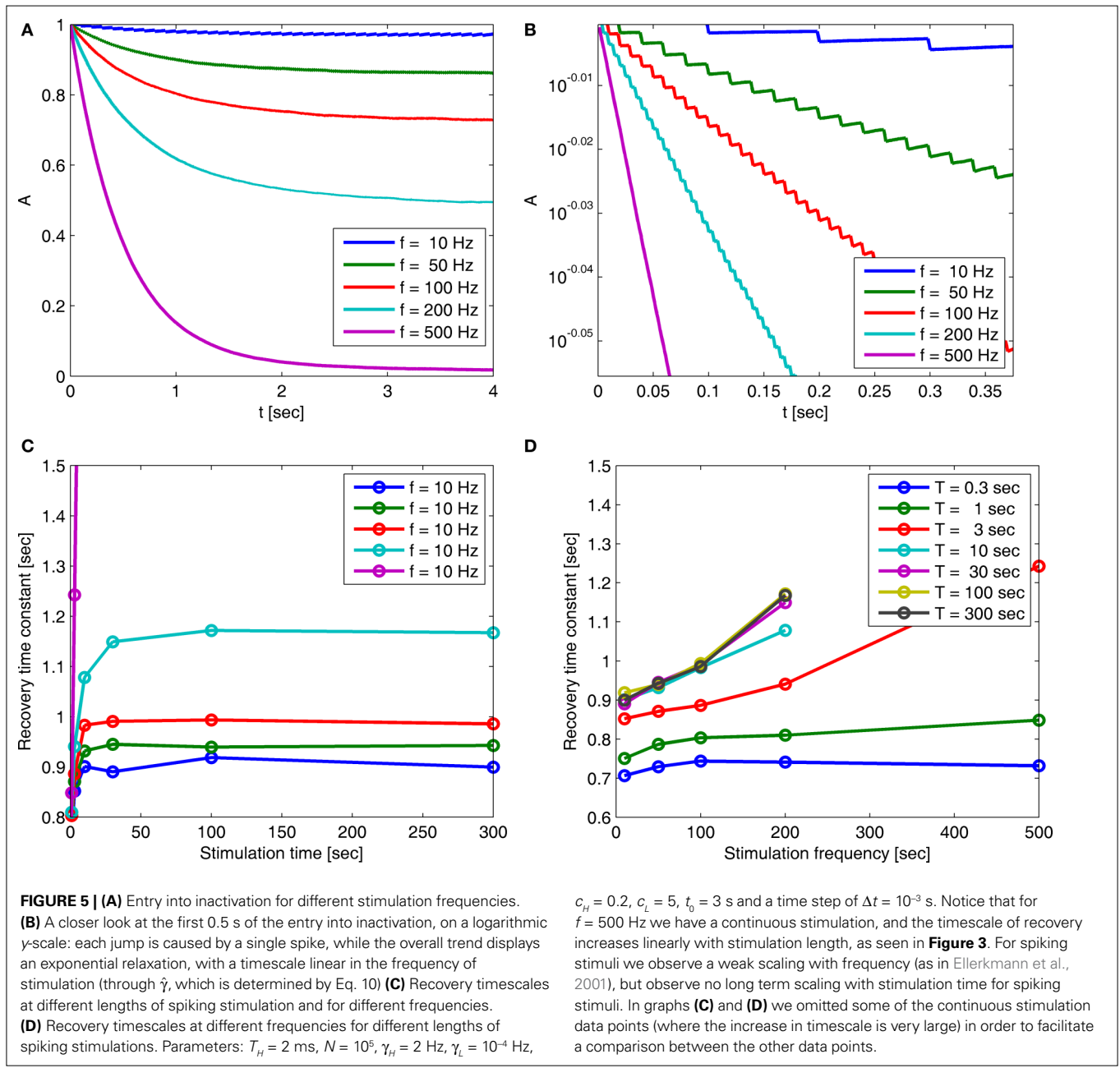

**FIGURE 5 | (A)** Entry into inactivation for different stimulation frequencies. **(B)** A closer look at the first 0.5 s of the entry into inactivation, on a logarithmic $y$-scale: each jump is caused by a single spike, while the overall trend displays an exponential relaxation, with a timescale linear in the frequency of stimulation (through $\hat{\gamma}$, which is determined by Eq. 10). **(C)** Recovery timescales at different lengths of spiking stimulation and for different frequencies. **(D)** Recovery timescales at different frequencies for different lengths of spiking stimulations. Parameters: $T_H = 2$ ms, $N = 10^6$, $\gamma_H = 2$ Hz, $\gamma_L = 10^{-4}$ Hz,

$c_H = 0.2$, $c_L = 5$, $t_0 = 3$ s and a time step of $\Delta t = 10^{-3}$ s. Notice that for $f = 500$ Hz we have a continuous stimulation, and the timescale of recovery increases linearly with stimulation length, as seen in **Figure 3**. For spiking stimuli we observe a weak scaling with frequency (as in Ellerkmann et al., 2001), but observe no long term scaling with stimulation time for spiking stimuli. In graphs **(C)** and **(D)** we omitted some of the continuous stimulation data points (where the increase in timescale is very large) in order to facilitate a comparison between the other data points.

drastically due to history-dependent effects. This may happen, for example, if $c(V_{rest}) > 3$, and the input to the neuron is sufficiently sparse (so that it does not cause continuous depolarizations for long times).

In Uebachs et al. (2006), a related stimulation experiment was performed on three types of calcium channels, each of which possesses different inactivation properties (which may be described in our model using different values of $\gamma$, $c$ and $t_0$). A slow-down of the timescale of recovery after a long −55 mV depolarizations was observed, similar to what was measured in sodium channels (referred to as 'continuous' in Figure 8 in Uebachs et al., 2006). In contrast, after a spiking stimulus at a maximal voltage of 25 mV with $T_H/(T_H + T_L) \approx 0.056$ ('mock APs' in Figure 8 in Uebachs et al., 2006), there was no observed increase in the timescale of recovery with the length of stimulation, in agreement with our model. Also, when the spiking stimulus was superimposed on the (continuous) step depolarization, the history-dependence of the timescale diminished ('continuous + mock APs' in Figure 8 in Uebachs et al., 2006). This may imply that $c(V)$ is not monotone in calcium channels – if $c(V)$ decreases between −95 and −55 mV and then increases between −55 and 25 mV, $\hat{c}$ should change less during the stimulus, and therefore the recovery timescale should also change less. This result might explain why Hering et al. (2004) did not observe any change in the recovery timescale after different lengths of depolarizations, given at the higher voltage of −20 mV.

## TEMPORAL CORRELATIONS

In Section 'The Joint Probability Distribution of the Availability', we present a simple approach to calculating the joint moments and distributions to be in the available state at different times. Specifically, we derive Eq. 57, the asymptotic behavior of the stationary auto-covariance of a single channel, for $c > 1$:

$$k(t) = \left[ p_\infty^2 \left( 1 - p_\infty \right) t_0^{c-1} \right] t^{-c+1}. \tag{11}$$

For $0 < c < 1$, $k(t)$ is not defined since in that case the channel is not stationary and decays to complete inactivation. Also, in the Supplementary Material we show that the auto-covariance function of the population availability under voltage clamp is simply $k(t)/N$, where $N$ is the size of the channel population.

Notice that for $1 < c < 2$, we get that $\int_0^\infty k(t)dt = \infty$. This implies that the process is characterized by a long-memory (see definition on page 42 of Beran, 1994). This result complies with the results of Mercik and Weron (2001) for potassium channels (on timescales <1 s), as also pointed out by Goychuk and Hanggi (2004).

## RELATION TO PREVIOUS WORK

Several previous studies have attempted to construct a mathematical modeling framework which can account for long term temporal correlations, power-law behavior, and history-dependence of responses related to ion channels. The initial work along these lines was motivated by single channel experiments on time scales of milliseconds-to-seconds (see Liebovitch and Sullivan, 1987; Millhauser et al., 1988; Marom, 1998 and the references therein), and therefore does not directly imply the tail behavior of the RTPDF on the timescale of seconds-to-minutes, which was explored in Toib et al. (1998). In this brief comparative discussion we focus on models addressing channel dynamics.

As far as we are aware, the paper by Millhauser et al. (1988) was the first to propose a microscopic framework for generating RTPDFs with power-law temporal dependence, based on a large set of inactivation Markovian states. This work constructed RTPDFs of the form $t^{-\alpha}$ for $1/2 \leq \alpha \leq 3/2$ which correspond to $-1/2 \leq c \leq 1/2$ in our model. As was mentioned in that work, $\alpha \leq 1$ (or $c \leq 0$) contradicts the normalization of the RTPDF, and therefore cannot be correct for long timescales. However, for intermediate time scales, similar to those observed in single channel experiments, such behavior may be possible. A main focus of that work was proving that when the transitions between these inactivation states resemble a diffusion process, it leads to a RTPDF of the form $t^{-3/2}$, which corresponds to $c = 1/2$. As stated in Section 'Relaxation to a steady state under a step voltage', such a value of $c$ always leads to complete inactivation of the channel in the long run, and therefore cannot be correct for longer timescales. All later models based on this diffusion model suffer from the same type of difficulty. A related line of work was pursued around the same time in Liebovitch and Sullivan (1987). This approach, as well as that of Lowen and Teich (1995), proposed several types of RTPDF, which reduce to a power law in special cases. However, we note that the main concern of Liebovitch and Sullivan (1987), Lowen and Teich (1995) and Millhauser et al. (1988) was the establishment of RTPDFs consistent with single channel experimental findings, rather than on providing an analysis of the long term channel dynamics that is the main focus of this paper.

A more recent line of work (Goychuk and Hanggi, 2004), leaning on earlier work in the statistical physics community, formulated channel dynamics in the form of the so-called generalized Master equation. This work allows for general asymptotic power-law dependence of the RTPDF, and was shown to lead to experimentally measured (Mercik and Weron, 2001) power-law decay of single channel correlations, as we obtain in Eq. 11. However, this work does not directly address the main experimental finding from Toib et al. (1998), namely an exponential recovery process with a history-dependent time scale. Conceptually, two main features distinguish this work from ours. First, input (voltage) dependent parameters, an essential feature of our work, cannot be dealt with through an approach based on the kernels used in Goychuk and Hanggi (2004) (defined through their Laplace transforms). Second, our methods additionally enable the explicit calculation of the channel probability to be available, and of its history-dependence and joint moments (to any order), in many important cases.

The model of Millhauser et al. (1988) was recently used in Gilboa et al. (2005) in order to directly explain the experimental results in Toib et al. (1998). They were able to show, through numerical simulation and approximate analytic solutions of an equivalent diffusion model, that multiple time scales history-dependent behavior (of the type described in Results) is indeed reproduced within the diffusion model. However, the exponential nature of the recovery was not addressed. More recently Friedlander and Brenner (2009) analyzed history-dependent phenomena in the case of RTPDFs with power-law behavior corresponding to $0 < c < 1$ in our model, using the framework developed in Goychuk and Hanggi (2004). Specifically, they focused on the mapping between the input (voltage) and the output (availability) for step inputs, demonstrating power-law recovery. Our work corroborates these results, and provides exact analytic expressions for the recovery rates. Moreover, the input–output view is extended to include arbitrary time-dependent inputs and input dependent recovery.

A different line of work was recently suggested in Marom (2009), whereby the complex history-dependent channel dynamics is reduced to a single local in time logistic like equation, where the recovery rate depends on the level of activation in a power-law fashion. This approach can be viewed as a zero order approximation of the full dynamics given in Eq. 43, whereby the entire history is replaced by a single reporter, which is a function of the current availability. More generally, one can envisage replacing the complete dynamics by a finite set of differential equations containing a truncated list of moments of the distribution of recovery time scales. Such an approach, while introducing a further approximation step, falls within the widely studied field of Markovian population dynamics, and offers, due to its mathematical simplicity, the potential of being smoothly incorporated into higher level models of single neurons, viewed as a population of channels.

Along similar lines, the work of Lowen et al. (1999) uses a population of semi-Markovian with power-law RTPDF, coupled to an equation describing the voltage dynamics, to numerically reproduce many of the long term behaviors of single neurons. It is quite reasonable to assume that the long-range correlations reported in Lowen et al. (1999) are the direct result of the power-law RTPDF with $1 < c < 2$, which was used for the ion channels in that paper.

## DISCUSSION

In this work we have thoroughly analyzed the dynamics of a generic two-state model of an ion channel consisting of a Markovian state and a non-Markovian state. We derived a general dynamic equation for the probability of the channel to be available, or equivalently, the fraction of available channels in a voltage clamped population. This dynamic equation, which is derived for both constant (Eq. 3) and time-dependent (Eq. 43) voltage, defines a direct input (voltage) to output (availability) relation. We derived explicit solutions to this equation in many important cases, and studied their properties. Specifically, we considered the asymptotic power-law approach of the channels to steady state under constant voltage (Eq. 4), the distribution of recovery timescales in the population following an abrupt change in voltage (Eq. 36), and the time-averaging of rapid fluctuations in the voltage (Eq. 8). Also, we derived simple expressions for all joint moments of the availability and specifically, for the auto-covariance function of the channel (Eq. 11). An interesting conclusion of this analysis is that the channel is characterized by four different 'modes' in which its behavior is rather different, depending on the value of $c$ (see **Table 1**).

The main experimental finding from Toib et al. (1998), an exponential recovery process with a history-dependent timescale, was fully reproduced. We constrained, for several channel types and different voltages, the model parameter space (voltage-dependent $c$ and $\gamma$ and constant $t_0$) by using Toib et al. (1998) as well as other channel population experiments (Ellerkmann et al., 2001; Hering et al., 2004; Uebachs et al., 2006). Also, we addressed the issue of long-memory phenomena (Mercik and Weron, 2001).

Our model presents many quantitative predictions on the dynamics of ion channel populations, for different kinds of voltage inputs, especially for long times. These predictions can be confirmed by experiments similar to Toib et al. (1998), but in which voltage stimuli are more variable – in both voltage values and temporal shape. Some of the specific qualitative predictions are the following. The value of the parameter $c$ at rest obeys $c(V_{rest}) > 1$ in all healthy channels (see Relaxation to a steady state under a step voltage). From Section 'History-dependent recovery timescale in response to voltage pulse' we conclude that the recovery from slow inactivation of Na$II$ and Na$IIA$ channels is voltage dependent, and for these channels, there exists some range of voltages for which $1 < c < 3$, where the recovery becomes non-exponential. Finally, we predict that sparse spiking stimuli (e.g., the neuron's own action potentials) can induce slow inactivation but only minor changes in the timescale of recovery from inactivation (see Channel response to a spiking stimulus).

One of the more interesting questions pertains to the relation between the history-dependent channel dynamics and the generation of action potentials in the cell, and, more concretely, the possible functionality of such behavior (Lundstrom et al., 2008; Wark et al., 2009). In order address this issue clearly and fully, two further steps must be taken.

Experimentally, it is necessary to find the correct model parameters (voltage-dependent $c$ and $\gamma$ and constant $t_0$) for different voltages and different types of channels. In particular, the value of $c$ at $V_{rest}$ and its average value during physiologically realistic cellular stimuli are critical for determining whether history-dependent relaxation and long term temporal correlations actually occur during normal cell activity. If, for example, $c > 3$ effectively for all channels in the cell, then we should not expect that these phenomena affect action potential generation – since in this case the inactivated state is approximately Markovian.

Theoretically, it is necessary to construct a model of a neuron that incorporates the type of channel studied here, as well as other types of channels which occur in the cell membrane. Using such a model we could determine the nature of the feedback interaction between channel activity and the membrane voltage, and the impact of this interaction on the action potential generation probability (e.g., Lowen et al., 1999; Gilboa et al., 2005). For example, it is quite reasonable to assume that the history-dependent relaxation exhibited at the single neuron level (Fairhall et al., 2001; Lundstrom et al., 2008; Wark et al., 2009) may be caused by the channel mechanism discussed here – especially since slow inactivation in sodium channels is known to have a strong effect on neuronal adaptation at these timescales (Fleidervish et al., 1996; Powers et al., 1999). If this is indeed the same mechanism described by our model, then replacing the continuous stimulation by spike stimulation should greatly reduce such behavior. The persistence of history-dependent relaxation in this setting, would imply that other processes (e.g., multiple interacting channel types) are in place. Another issue that can be explored using this method, is whether possible long term temporal correlations in the channel are the cause of similar long-memory phenomena at the cellular level (Soen and Braun, 2000). In fact, the work of Lowen et al. (1999) argues persuasively along these lines, as many of the long-range temporal behaviors of neurons are well replicated within a simple model for the membrane potential incorporating non-Markovian channel dynamics. Indeed, extending the mathematical tractability of the present approach to the higher level of a neuron is a major theoretical challenge.

Finally, it is important to observe that most of the mathematical results presented in this work are general, and can be extended to other models in which the RTPDF of the unavailable state is not a power law. The mathematical framework that was developed here to model ion channels is quite flexible, and may be used to describe other systems in which one can similarly define separate available and unavailable states so that the transitions between these states are random and depend only on the residence time of the state and the external input (e.g., voltage or ligand concentration). For example, inactivation of cellular receptors could be modeled similarly, as pointed out by Friedlander and Brenner (2009). The method for mapping a general Markov model to our two-state non-Markov model is described in detail in Section 'How to create a two-state non-Markov model from a general Markov model'.

**Table 1 | Modes of behavior for different ranges of $c$.**

| $c$ Range | Mode of behavior |
|---|---|
| $(0, 1)$ | The channel is non-stationary and decays to complete inactivation. Recovery timescales increase linearly with a constant dispersion. |
| $(1, 2)$ | The channel has a stationary steady state in which it is partially available, and recovery is non-exponential. The channel auto-correlation function possesses long memory. |
| $(2, 3)$ | The channel has a stationary steady state in which it is partially available and recovery is non-exponential. |
| $(3, \infty)$ | The channel has a stationary steady state in which it is partially available, the recovery timescale distribution has a finite mean and dispersion and is near-exponential. The inactivated state is 'almost Markovian'. |

## METHODS

In this section we establish several general analytical methods, which can be used in any two-state model in which one state is Markovian while the other state is non-Markovian, namely $\psi_g(t)$ is still exponential as in Eq. 1, but $\psi_p(t)$ may be replaced by some other general RTPDF, which we denote by $\psi_f(t)$. When establishing general results we use $\psi_f(t)$, and when specializing them to the power-law RTPDF we use $\psi_p(t)$.

We present two different formalisms – the Markovian-process based 'projection method' in Section "'Projection' of the Inactivated State" and the 'Renewal Theory' approach in Section 'Renewal Theory Approach'. This was done in order to explore different aspects of the channel dynamics in the case of constant voltage input. Moreover, we show how to map an existing Markov model to a two-state non-Markov model. In Section 'General Voltage Input' we use the presented methods to derive the dynamics in the case of general input. In Section 'Oscillating Voltage Input' we consider the case of rapidly oscillating input. Finally, in Section 'The Joint Probability Distribution of the Availability' we calculate the joint distributions of the channel.

### 'PROJECTION' OF THE INACTIVATED STATE

In this section we assume that the voltage input is constant, so that all the model parameters are also constant. We wish to make use of the well-established formalism of Markov processes to derive the dynamics of our non-Markovian model. To do so we needed to find a way to replace the non-Markovian inactivated state with an equivalent Markov inactivation state space. For example, in the context of the specific channel model with the power-law RTPDF $\psi_p(t)$, there are many possible state spaces that may be used to produce it (Liebovitch, 1989). Some of them may be physiologically more accurate than others, but, mathematically, this is inconsequential to the channel dynamics. And so, we chose the simplest state space, in which all inactivation states are parallel – each state being connected only to the available state, as seen in **Figure 6**. Though the figure depicts a finite set of states, in the actual model we use a continuum of states, so that the equivalence between the models is exact. We use as the continuous 'index' of each activation state its recovery rate, $\delta$. Since it is Markovian, the RTPDF function of each inactivation state $\delta$ is exponential,

$$\psi_\delta(t) = \delta \exp(-\delta t), \quad t \geq 0 \ \ (0 \leq \delta < \infty).$$

Each time the channel is inactivated, it goes into one of the inactivation states. We denote by $f(\delta)$ the probability density function to go from the available state into a specific inactivation state $\delta$. This means that the inactivation rate from the available state into the inactivated states $(\delta, \delta + d\delta)$ is $\gamma f(\delta) d\delta$, and so the total rate of inactivation is $\int_0^\infty \gamma f(\delta) d\delta = \gamma$, as required. With this condition fulfilled this multiple-state Markovian model is equivalent to our two state non-Markovian model if we ensure that the RTPDF of the aggregation of all the inactivated states is $\psi_f(t)$. To fulfill this condition, we use law of total probability and demand that:

$$\psi_f(t) = \int_0^\infty f(\delta) \psi_\delta(t) d\delta. \tag{12}$$

### Dynamic equations

In order to derive the dynamical equations of the non-Markovian model, we begin with the Markovian model described in **Figure 6**, present its underlying equations, and then take the continuum limit.

We denote $\bar{\pi} = (p, \pi_1, \ldots, \pi_N)$, where $p$ is probability of being in the available state, and $\pi_k$ is the probability of being in the inactivated state $I_k$. The dynamic equation for the probability of this homogeneous Markov process is

$$\frac{d}{dt} \bar{\pi}(t) = \bar{\pi}(t) \mathbf{Q},$$

where

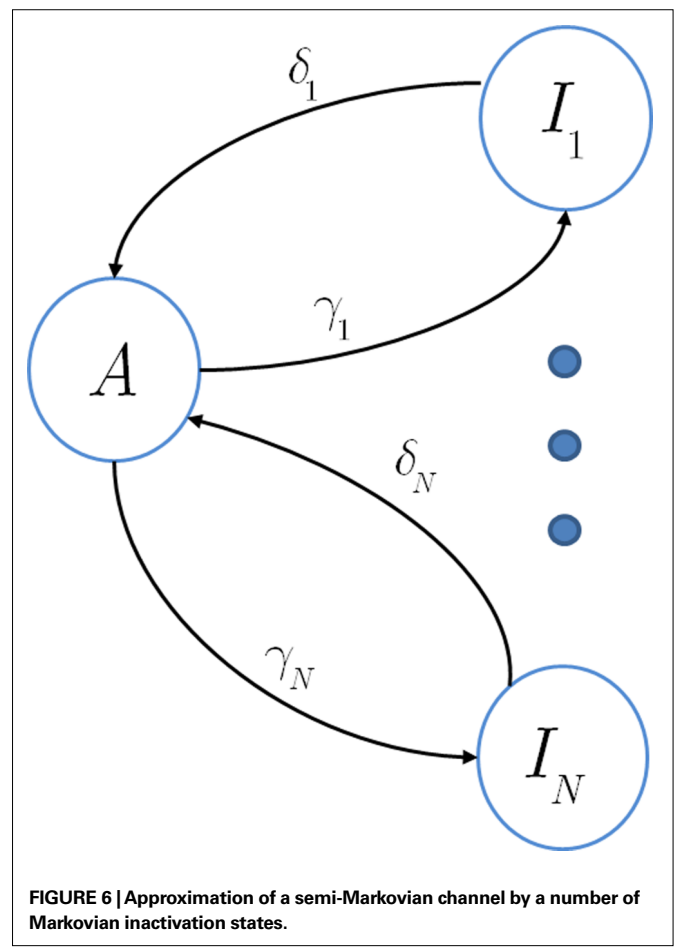

**FIGURE 6 | Approximation of a semi-Markovian channel by a number of Markovian inactivation states.**

$$Q = \begin{pmatrix} -\sum_{k=1}^{N} \gamma_k & \gamma_1 & \gamma_2 & \cdots & \gamma_N \\ \delta_1 & -\delta_1 & 0 & 0 & 0 \\ \delta_2 & 0 & -\delta_2 & 0 & 0 \\ \vdots & 0 & 0 & \ddots & 0 \\ \delta_N & 0 & 0 & 0 & -\delta_N \end{pmatrix},$$

is the rate matrix.

Writing this explicitly, we get

$$\frac{d}{dt}\pi_k(t) = \gamma_k p(t) - \delta_k \pi_k(t), \quad k = 1,2,\ldots,N$$

$$\frac{d}{dt}p(t) = -\sum_{k=1}^{N} \gamma_k p(t) + \sum_{k=1}^{N} \pi_k(t)\delta_k$$

In the exact continuum limit we substitute $\gamma_k \to \gamma f(\delta)d\delta$, $\pi_k(t) \to \pi_\delta(t)d\delta$, $\sum_{k=1}^{N} \to \int_0^\infty$, yielding:

$$\frac{d}{dt}\pi_\delta(t)d\delta = p(t)\gamma f(\delta)d\delta - \delta\pi_\delta(t)d\delta,$$

$$\frac{d}{dt}p(t) = -\gamma p(t) + \int_0^\infty \pi_\delta(t)\delta d\delta. \tag{13}$$

Dividing the first equation by $d\delta$ we obtain:

$$\frac{d}{dt}\pi_\delta(t) = \gamma f(\delta)p(t) - \delta\pi_\delta(t) \tag{14}$$

with solution

$$\pi_\delta(t) = \pi_\delta(0)e^{-\delta t} + \gamma f(\delta)\int_0^t p(t')e^{-\delta(t-t')}dt' \tag{15}$$

Inserting this into Eq. 13 and switching the order of integration we get:

$$\frac{d}{dt}p(t) = -\gamma p(t) + \int_0^\infty \pi_\delta(0)e^{-\delta t}\delta d\delta + \gamma \int_0^t dt'p(t')\int_0^\infty f(\delta)e^{-\delta(t-t')}\delta d\delta$$

Using Eq. 12, the expression can be re-written as:

$$\frac{d}{dt}p(t) = -\gamma p(t) + \gamma \int_0^t p(t')\psi_1(t-t')dt' + \int_0^\infty \pi_\delta(0)\psi_\delta(t)d\delta \tag{16}$$

This result is quite intuitive:

1. The term $-\gamma p(t)$ represents the probability current out of the available state.
2. The term $\gamma \int_0^t p(t')\psi_1(t-t')dt'$ represents the probability current that goes from availability to inactivation and back again – where we sum over all the possibilities to inactivate at time $t'$ and then recover at time $t$.
3. The term $\int_0^\infty \pi_\delta(0)\psi_\delta(t)d\delta$ represents the probability current from the initial inactivation states – the probability to be at each inactivation state at time $t = 0$, times the probability density to recover exactly at time $t$.

Such equations are commonly dealt with in the Laplace domain. Using a Laplace transform on Eq. 16, we get:

$$s\tilde{p}(s) - p(0) = -\gamma\tilde{p}(s) + \gamma\tilde{p}(s)\tilde{\psi}_1(s) + \int_0^\infty \frac{\delta\pi_\delta(0)}{s+\delta}d\delta,$$

$$\tilde{p}(s) = \frac{p(0) + \int_0^\infty \frac{\delta\pi_\delta(0)}{s+\delta}d\delta}{s + \gamma\left(1 - \tilde{\psi}_1(s)\right)}, \tag{17}$$

where the tilde denotes the Laplace transform. The final term in the numerator results from the initial probability distribution in the inactivated states. If at $t = 0$ the channel is in the available state, namely $p(0) = 1$, then this term vanishes.

### Projection of a power-law RTPDF

By using an inverse Laplace transform on the $t$ variable in Eq. 12, we can easily show that in the case where $\psi_1(t) = \psi_p(t)$, $f(\delta)$ is distributed according to the gamma distribution,

$$f(\delta) = \frac{t_0^c}{\Gamma(c)}\delta^{c-1}\exp(-\delta t_0), \quad \delta > 0, \tag{18}$$

where $\Gamma(\cdot)$ is the gamma function. Thus, the aggregation of all the inactivation states has the RTPDF:

$$\int_0^\infty f(\delta)\psi_\delta(t)d\delta = \frac{t_0^c}{\Gamma(c)}\int_0^\infty \delta^c \exp\left(-\delta(t+t_0)\right)d\delta$$

which equals $\psi_p(t)$ as required.

### Asymptotic solution of the dynamic equations for power-law RTPDF

In this section we demonstrate how to derive an asymptotic solution of Eq. 17 in the case where $\psi_1(t) = \psi_p(t)$. We assume throughout this section that $c$ is non-integer; the limiting case where $c$ is an integer is discussed in the Supplementary Material. In order to understand the behavior of $p(t)$ for large values of $t$, we need to consider the small $s$ limit of $\tilde{p}(s)$. To do so, we first calculate $\tilde{\psi}_p(s)$. In the Supplementary Material we prove that for non-integer values of $c$,

$$\tilde{\psi}_p(s) = ce^{st_0}\left[\Gamma(-c)\cdot(t_0 s)^c - \sum_{k=0}^\infty \frac{(-st_0)^k}{k!(k-c)}\right] \tag{19}$$

Retaining the leading order in $s$, we have (see Supplementary Material):

$$\tilde{\psi}_p(s) = 1 + \left(\frac{t_0}{1-c}\right)s + \frac{t_0^2}{(c-2)(c-1)}s^2 + c\Gamma(-c)t_0^c \cdot s^c + O\left(s^{\min(3,c+1)}\right) \tag{20}$$

and substituting Eq. 20 into Eq. 17, we find:

$$\tilde{p}(s) = \frac{1}{s - \frac{\gamma t_0}{1-c}s - c\gamma\Gamma(-c)t_0^c s^c + O\left(s^{\min(2,c+1)}\right)}.$$

If $c < 1$,

$$\tilde{p}(s) = \frac{-1}{c\gamma\Gamma(-c)t_0^c}s^{-c} + O\left(s^{1-2c}\right). \tag{21}$$

while if $c > 1$,

$$\tilde{p}(s) = p_\infty \cdot s^{-1} + c\gamma\Gamma(-c)p_\infty^2 t_0^c s^{c-2} + O(1) \qquad (22)$$

where we have introduced

$$p_\infty = \begin{cases} \frac{-c-1}{\gamma t_0 + c - 1} & \text{if} \quad c \geq 1, \\ 0 & \text{if} \quad 0 < c < 1, \end{cases} \qquad (23)$$

denoting the steady state value of $p(t)$ – a result that can be confirmed by using the final value theorem on $\tilde{p}(s)$. Also, Notice that in Eq. 22 we deliberately retained the $s^{c-2}$ term, even though $s^{c-2} = O(1)$ for $c > 2$. The reason for this will become clear soon.

We wish to find the asymptotic behavior of $p(t) = \mathcal{L}^{-1}[\tilde{p}(s)]$ as $t \to \infty$. To do this we use a Tauberian theorem (Theorem 4 on page 446 in Feller, 1971), stating that $f(t) \sim [K/\Gamma(\alpha)]t^{\alpha-1}$ as $t \to \infty$ if, and only if, $\mathcal{L}[f(t)](s) \sim Ks^{-\alpha}$ as $s \to 0$, where $K$ is some constant and $\alpha > 0$.

For $0 < c < 1$, a direct application of the theorem to Eq. 21 gives:

$$p(t) = \frac{-1}{c\gamma\Gamma(-c)\Gamma(c)t_0^c}t^{c-1} = \frac{\sin(\pi c)}{\gamma\pi t_0^c}t^{c-1},$$

where we used the result $\Gamma(c)\Gamma(-c) = -\pi/c\sin(\pi c)$ (page 3 in Erdeyi, 1953, vol. I).

For $1 < c < 2$, we first subtract from Eq. 22 the Laplace transform of the steady state value, $p_\infty s^{-1}$, and then use the Tauberian theorem. From the linearity of the Laplace transform we get:

$$p(t) - p^* = c\gamma\frac{\Gamma(-c)}{\Gamma(2-c)}p_\infty^2 t_0^c t^{1-c},$$

$$= (1 - p_\infty)p_\infty t_0^{c-1} t^{1-c}$$

where we used the result $\Gamma(-c)/\Gamma(2 - c) = -1/(c(1 - c))$ (page 3 in Erdeyi, 1953, vol. I), and Eq. 23.

For $c > 2$, we can again subtract the steady state term $p_\infty s^{-1}$ from Eq. 22, but now we are left with only positive powers of $s$, and therefore we cannot directly apply the Tauberian theorem. To circumvent this problem, we define $m = \lceil c - 2 \rceil$, the upper integer part of $c - 2$, and differentiate $m$ times:

$$\frac{d^m}{ds^m}(\tilde{p}(s) - p_\infty s^{-1}) = c\gamma\Gamma(-c)p_\infty^2 t_0^c \frac{\Gamma(c-1)}{\Gamma(c-1-m)}s^{c-2-m} + O(1)$$

Since that all the integer order terms with degree lower than $m$ vanish after we differentiate for $m$ time. Using the fact that:

$$\mathcal{L}[t^m f(t)] = (-1)^m \frac{d^m}{ds^m}\mathcal{L}[f(t)],$$

the linearity of the Laplace transform and the above Tauberian theorem, we get:

$$(-1)^m t^m \cdot (p(t) - p_\infty) = c\gamma p_\infty^2 \frac{\Gamma(-c)\Gamma(c-1)}{\Gamma(c-1-m)\Gamma(m+2-c)}t_0^c t^{1-c+m},$$

$$p(t) - p_\infty = (1 - p_\infty)p_\infty \frac{(-1)^m \sin((c-2-m)\pi)}{\sin(c\pi)}t_0^{c-1} t^{1-c}$$

$$= (1 - p_\infty)p_\infty t_0^{c-1} t^{1-c}$$

where we used similar identities for $\Gamma(\cdot)$ as used previously, Eq. 23 and the fact that $m$ is an integer. Notice that we can use this method to find higher orders of the asymptotic behavior of $p(t)$ from the higher fractional order terms in $s$.

An interesting implication of the above derivation is that the asymptotic behavior of $f(t)$ is completely determined by the minimal fractional power of $s$ in $\mathcal{L}[f(t)]$ (assuming that $\mathcal{L}[f(t)]$ can be uniquely represented as a countable sum of integer and fractional powers of $s$). Because of this, any (minimal) linear time invariant system will behave asymptotically as a power law if it contains some component with a power-law kernel – which always introduces a fractional power of $s$ into the system's transfer function in the Laplace domain.

In conclusion, we found the asymptotic behavior of $p(t)$ as $t \to \infty$, for all $c$,

$$p(t) = \begin{cases} p_\infty + (1 - p_\infty)p_\infty t_0^{c-1} t^{1-c} & \text{if} \quad 1 < c \\ \frac{\sin(\pi c)}{\pi p_0^c}t^{c-1} & \text{if} \quad 0 < c < 1 \end{cases} \qquad (24)$$

We derived this result for the case $p(0) = 1$, but it remains valid also when $p(0) \neq 1$ if the expression $\int_0^\infty [\delta\pi_\delta(0)/(s + \delta)]d\delta$ is analytic at the origin. For example, this always is true if for all $\delta < \delta_0$, $\pi_\delta(0) = 0$ for some $\delta_0 > 0$ – meaning that there exists a finite supremum to the recovery timescales of all the inactivation states in the initial distribution. There is another way to see that the asymptotic solution in Eq. 24 is not sensitive to the initial conditions. Suppose we set $p(t_i) = 1$ for some $t_i$, instead of $p(0) = 1$. The asymptotic solution is $p(t - t_i)$. Since

$$(t - t_i)^a = t^a \left(1 - \frac{t_i}{t}\right)^a = t^a \left(1 - \frac{at_i}{t} + O(t^{-2})\right)$$

$$= t^a + O(t^{a-1}),$$

and therefore Eq. 24 is changed by the addition of an asymptotically negligible term. This asymptotic insensitivity to initial conditions stands in contrast with the case where the inactivated state is Markovian, and the solution is exponential,

$$p(t) = p_\infty + (p(0) - p_\infty)e^{-at}.$$

In this case when we set the initial condition to $t_i$, instead of 0, we get,

$$p(t) = p_\infty + (p(t_i) - p_\infty)e^{-a(t-t_i)}$$

$$= p_\infty + e^{at_i}(p(0) - p_\infty)e^{-at}.$$

Therefore, the solution is changed by a pre-factor, which is not asymptotically negligible as in the power-law case.

### How to create a two-state non-Markov model from a general Markov model

So far, we used the Markov process formalism to find a dynamic equation for a non-Markov two state model. Now consider the converse problem: suppose we have an existing Markov model for an ion channel, how can we transform it to a two-state non-Markov model?

To answer this question we use a method based on the ideas presented in Colquhoun and Hawkes (1977). Recall that the dynamics of a homogeneous Markov process are governed by the equation $d\bar{\pi}(t)/dt = \bar{\pi}(t)\mathbf{Q}$, where $\bar{\pi}(t)$ is the probability vector to be at each state, and $\mathbf{Q}$ is the rate Matrix. We divide the Markov state space into two disjoint sets of states, according to their conductivity – $A$, the available states and $I$, the inactivated states. These two sets will be the two states in the non-Markov model. We divide the probability vector into two vectors according to these sets $\bar{\pi}(t) = (\bar{\pi}_A(t), \bar{\pi}_I(t))$. Similarly, we divide the rate matrix $\mathbf{Q}$ into block matrices corresponding to these sets of states:

$$\mathbf{Q} = \begin{pmatrix} \mathbf{Q}_{AA} & \mathbf{Q}_{AI} \\ \mathbf{Q}_{IA} & \mathbf{Q}_{II} \end{pmatrix}.$$

In order to calculate the RTPDFs of the inactivated states, we introduce a modified Markov process with probability vector $\tilde{\pi}(t) = (\tilde{\pi}_A(t), \tilde{\pi}_I(t))$, and the rate matrix:

$$\tilde{\mathbf{Q}} = \begin{pmatrix} 0 & 0 \\ \mathbf{Q}_{IA} & \mathbf{Q}_{II} \end{pmatrix}.$$

The dynamics of the modified process are similar to the original process, except that now the available states have become absorbing – each time the process reaches these states, it remains there forever. From $d\tilde{\pi}(t)/dt = \tilde{\pi}(t)\tilde{\mathbf{Q}}$ we get,

$$\frac{d}{dt}\tilde{\pi}_I(t) = \tilde{\pi}_I(t)\mathbf{Q}_{II} \tag{25}$$

$$\frac{d}{dt}\tilde{\pi}_A(t) = \tilde{\pi}_I(t)\mathbf{Q}_{IA} \tag{26}$$

The solution of Eq. 25 can be written in the form,

$$\tilde{\pi}_I(t) = \tilde{\pi}_I(0)\exp(\mathbf{Q}_{II}t), \, t \geq 0, \tag{27}$$

where the exponent of a matrix $\mathbf{A}$ is defined as $\exp(\mathbf{A}) = \sum_{k=0}^{\infty} \mathbf{A}^k/k!$.

Now assume that $\bar{\pi}(0) = \tilde{\pi}(0) = (0, \bar{\pi}_I(0))$, meaning that both processes at $t = 0$ are distributed identically, over the inactivated states. The probability that the original process has jumped to some available state before time $t$ is equal to probability that the modified process is in some available state at time $t$. This probability can be written as $\tilde{\pi}_A(t)\bar{u}_A^T$, where we defined $\bar{u}_A^T = (1,1,\ldots,1)^T$, a column vector of ones with the same number of components as $\tilde{\pi}_A(t)$. Differentiating this expression, we get the RTPDF of the inactivated state, assuming that the channel has inactivated at $t = 0$:

$$\psi_I(t) = \frac{d}{dt}\left(\tilde{\pi}_A(t)\bar{u}_A^T\right) = \bar{\pi}_I(0)\exp(\mathbf{Q}_{II}t)\mathbf{Q}_{IA}\bar{u}_A^T, \quad t \geq 0, \tag{28}$$

where we used Eqs 26 and 27, and $\bar{\pi}_I(0)$ represents the probability distribution in the inactivated states immediately after the inactivation. Similarly, the RTPDF of the available state is

$$\psi_A(t) = \bar{\pi}_A(0)\exp(\mathbf{Q}_{AA}t)\mathbf{Q}_{AI}\bar{u}_I^T, \quad t \geq 0. \tag{29}$$

where $\bar{u}_I^T$ is defined similarly to $\bar{u}_A^T$.

Note that $\psi_I(t)$, $\psi_A(t)$ are always a sum of exponents, but after the transformation to the two state model is complete, we can approximate them by some simpler functional form. The form of these RTPDFs (in Eqs 28 and 29) pose a problem, through their dependence on $\bar{\pi}_A(0)$ and $\bar{\pi}_I(0)$. Assume that the channel inactivated at time $t_1$, and recovered at time $t_2$. Generally, $\bar{\pi}_I(t_1)$ may depend on the time that the channel was available before $t_1$, and $t_2 - t_1$ may depend on $\bar{\pi}_I(t_1)$. Therefore, the time the channel spends in inactivation may depend on the previous duration of the channel in the available state, which may depend on the previous duration the channel in the inactivated state, and so on. This would prevent us from constructing a model with two *independent* non-Markov states.

There are some specific cases when this problem vanishes. One of them is when $A$ can be approximated as single Markov state. This happens, as we mentioned in Section 'Background', when the rates between the available states are much larger than the rates of inactivation. In this case, as we show in the Supplementary Material,

$$\psi_A(t) = \psi_E(t) = \gamma \exp(-\gamma t), \quad t \geq 0, \tag{30}$$

$$\psi_I(t) = \bar{\pi}_A^s \mathbf{G}_{AI}\exp(\mathbf{Q}_{II}t)\mathbf{Q}_{IA}\bar{u}_A^T, \quad t \geq 0, \tag{31}$$

where we defined $\gamma = \bar{\pi}_A^s \mathbf{Q}_{AI}\bar{u}_I^T$, $(\mathbf{G}_{AI})_{mn} = (\mathbf{Q}_{AI})_{mn}/[\sum_{l=1}^{L}(\mathbf{Q}_{AI})_{nl}]$ and $\bar{\pi}_A^s$ is the stationary distribution in the available set of states, when all the transitions to inactivation are forbidden. This is the two-state model which we aimed for: a Markovian available state, and a non-Markovian inactivated state.

## RENEWAL THEORY APPROACH

Despite its merits, the use of the projection method may be limited in certain cases. The problem lies in the use of $\pi_\phi(t)$, the probability distribution in the inactivation states. Knowledge of the probability to be in an 'inactivation state' implies that we have some additional information about the future recovery time of the channel. But, when a channel is inactivated, we know only for how long it has been inactivated, and not to which 'inactivation state' it has arrived. Since these inactivation states are not observable, this approach restricts the practical use of the initial conditions $\pi_\phi(0)$ in Eq. 16. And so, in order to quantify the history-dependent behavior of the channel in terms of observable quantities, we turn to a different formalism that uses only the observable $T$, which is the time during which the channel has already been inactivated. In this section, as the in the last section, we again assume that the voltage input is constant, so all the model parameters are also constant.

For a channel that has been inactivated for time $T$, the probability of recovery in the next time interval $\Delta t$ is

$$P\{t_R < T + \Delta t \,|\, t_R > T\} = \frac{P\{T < t_R < T + \Delta t\}}{P(t_R > T)},$$
$$= \frac{\psi_I(T)\Delta t + O(\Delta t^2)}{\int_T^\infty \psi_I(z)dz}, \tag{32}$$

where $t_R$ denotes the time it takes the channel to recover since its last inactivation.

Therefore, the rate of recovery immediately after time $T$ is

$$\lambda \triangleq \lim_{\Delta t \to 0} \frac{1}{\Delta t} P\left(t_R < T + \Delta t \,|\, t_R > T\right)$$
$$= -\frac{d}{dT} \log\left(\int_T^\infty \psi_I(z)dz\right), \quad (T \geq 0) \tag{33}$$

and the time-dependent timescale of recovery is its inverse $\tau \triangleq 1/\lambda$.

Next, we compute $f_t(T)$, the probability density function of $T$ at time $t$, for a single channel, which well approximates the distribution of $T$ throughout a sufficiently large population. We did so using a 'Renewal Theory' formalism, which is described in Cox (1962). In renewal theory language $T$ is called the 'backward-recurrence time' until the last inactivation event.

First, we denote by $k_n(t)$ the probability density function of the time of the $n$-th inactivation event, assuming that at $t = 0$ the channel was available ($p(0) = 1$). Clearly $k_n(t)$ is the probability density function of a random variable which is the sum of $2n - 1$ independent random variables – corresponding to all the residence times in the available state, $\{X_m\}_{m=1}^n$, and in the inactivated state, $\{Y_m\}_{m=1}^{n-1}$, that occurred until the $n$-th inactivation event. Since the RTPDF of $X_m, Y_m$ are, respectively $\psi_E(t)$ and $\psi_I(t)$, and the probability density function of a sum of independent random variables is a convolution of the probability density functions of all the random variables, we get,

$$k_n(t) = \psi_E(t) * \underbrace{\left(\psi_I(t) * \psi_E(t)\right) * \left(\psi_I(t) * \psi_E(t)\right) * \cdots}_{n-1 \text{ times}}$$

In the Laplace domain, this translates to,

$$\tilde{k}_n(s) \triangleq \mathcal{L}\left[k_n(t)\right] = \left(\frac{\gamma}{\gamma + s}\right)\left(\tilde{\psi}_I(s) \cdot \frac{\gamma}{\gamma + s}\right)^{n-1}. \tag{34}$$

Defining the rate of inactivation events:

$$h(t) \triangleq \lim_{\Delta t \to 0} \frac{1}{\Delta t} P \quad \text{(some inactivation event occured in } (t, t + \Delta t)\text{)},$$

we have $h(t) = \sum_{n=1}^\infty k_n(t)$ and in the Laplace domain:

$$\tilde{h}(s) = \sum_{n=1}^\infty \tilde{k}_n(s) = \frac{\gamma}{s + \gamma\left(1 - \tilde{\psi}_I(s)\right)}$$

where we used the geometric series summation formula. Notice that $\tilde{h}(s) = \gamma\tilde{p}(s)$ and so $h(t) = \gamma p(t)$, which is not surprising, since the rate of inactivation events is expected to be the probability to be in the available state, times the rate of inactivation in that state. We can now write an expression for $f_t(T)$, conditioned on the fact that the channel is inactivated at time $t$. For a channel at time $t$ to be inactivated for the last $T$ seconds, there must be an inactivation event at time $t - T$ and since that time the channel must not recover. Dividing this by $(1 - p(t))$, the probability that the channel is indeed inactivated, we get,

$$f_t(T) = (1 - p(t))^{-1} h(t - T) \int_T^\infty \psi_I(u)du, \quad (0 \leq T \leq t)$$

$$= (1 - p(t))^{-1} \gamma p(t - T) \int_T^\infty \psi_I(u)du, \quad (0 \leq T \leq t)$$

### The case of power-law RTPDF

In the context of the specific model with power-law RTPDF the rate of recovery immediately after time $T$ is

$$\lambda = -\frac{d}{dT} \log\left(\int_T^\infty \psi_P(u)du\right) = \frac{c}{T + t_0}, \quad (T \geq 0)$$

and the time-dependent timescale of recovery is its inverse:

$$\tau \triangleq \frac{1}{\lambda} = \frac{T + t_0}{c}, \quad (T \geq 0). \tag{35}$$

The distribution of $T$ in this case $\psi_I(t) = \psi_P(t)$ is $f_t(T) = (1 - 9(t))^{-1}\gamma p(t - T)(1 + T/t_0)^{-c}$, $(0 \leq T \leq t)$.

Because of the multiplication in the last expression, it is generally difficult to write an expression for $f_t(T)$ in the Laplace domain. We therefore examine $f_t(T)$ in the limit $t \to \infty$. We get, by using the same asymptotic derivation we used in deriving Eq. 24, that:

$$f_t(T) = \begin{cases} \frac{c-1}{t_0}\left(1 + \frac{T}{t_0}\right)^{-c} & \text{if} \quad 1 < c \\ \frac{\sin(\pi c)}{\pi t_0^c}(t - T)^{c-1}\left(1 + \frac{T}{t_0}\right)^{-c} & \text{if} \quad 0 < c < 1 \end{cases}, \quad (0 \leq T \leq t) \tag{36}$$

Note that for $0 < c < 1$, for any fixed value of $T$, the function $f_t(T)$ vanishes for large values of $t$. This implies that the probability distribution shifts to infinity as $t$ increases; see **Figure 7**.

Calculating the first two moments of $T$ in each case, defining $\langle T \rangle$ as the mean of $T$ and $\sigma_T$ its standard deviation we obtain the following results by series expansion (see Supplementary Material).

$$0 < c < 1: \begin{cases} \langle T \rangle = (1 - c)t + O(1) \\ \sigma_T = \sqrt{\frac{c(1-c)}{2}}\, t + O(1) \end{cases} \tag{37}$$

$$1 < c < 2: \begin{cases} \langle T \rangle = \frac{c-1}{3-c}t_0^{c-1}t^{2-c} + O(1) \\ \sigma_T = \sqrt{\frac{c-1}{3-c}}\, t_0^{(c-1)/2}t^{(3-c)/2} + O\left(t^{2-c}\right) \end{cases} \tag{38}$$

$$2 < c < 3: \begin{cases} \langle T \rangle = \frac{t_0}{c-2} + O\left(t^{2-c}\right) \\ \sigma_T = \sqrt{\frac{1-c}{c-3}}\, t_0^{(c-1)/2} \cdot t^{(3-c)/2} + O(1) \end{cases} \tag{39}$$

$$3 < c: \begin{cases} \langle T \rangle = \frac{t_0}{c-2} + O\left(t^{2-c}\right) \\ \sigma_T = \frac{t_0}{c-2}\sqrt{\frac{c-1}{c-3}} + O\left(t^{(3-c)/2}\right) \end{cases} \tag{40}$$

We verified numerically that all of the results for these moments and for $f_t(T)$ itself are valid for the relevant timescales, as is demonstrated in **Figures S1 and S2** and in Supplementary Material.

### GENERAL VOLTAGE INPUT

In the previous sections we used two different formalisms to analyze our model. One was the *Projection Method*, and in the other was the formalism of *Renewal Theory*. So far, these two approaches may seem unrelated, and dealt only with the case of constant parameters. In this section we shall use elements from both formalisms to derive a dynamic equation for $p(t)$ for any time-varying parameters. For example, in the context of the spe-

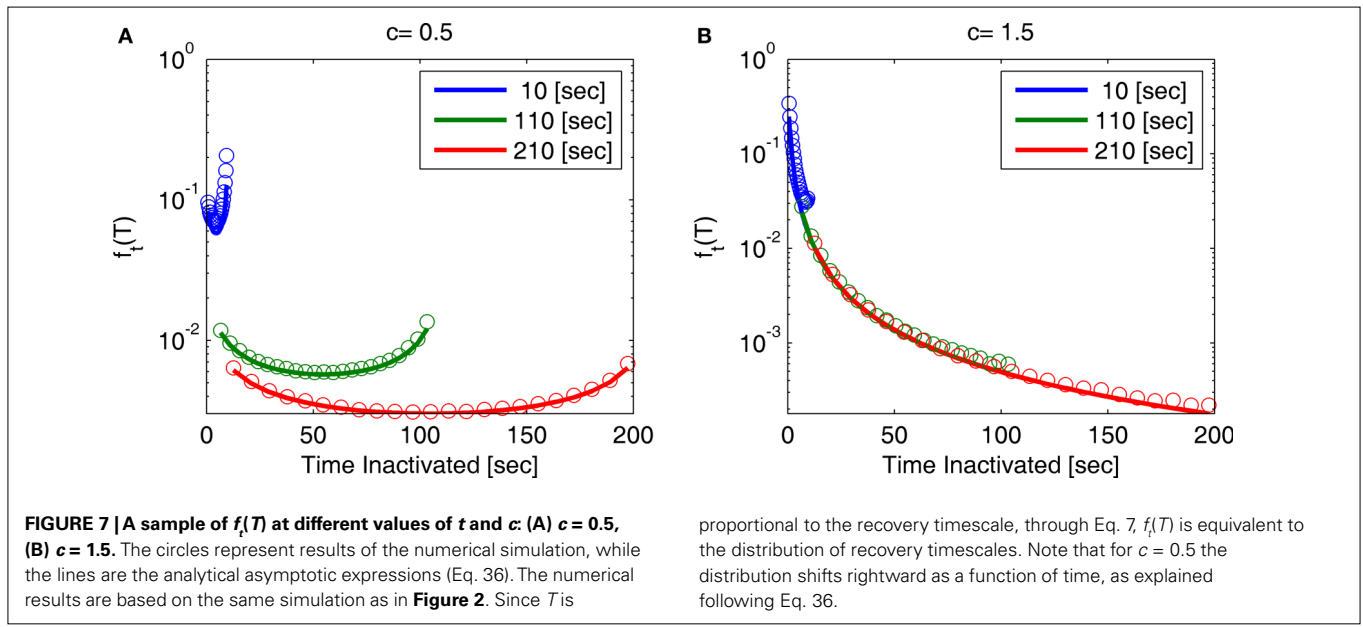

**FIGURE 7 | A sample of $f_t(T)$ at different values of $t$ and $c$: (A) $c = 0.5$, (B) $c = 1.5$.** The circles represent results of the numerical simulation, while the lines are the analytical asymptotic expressions (Eq. 36). The numerical results are based on the same simulation as in **Figure 2**. Since $T$ is

proportional to the recovery timescale, through Eq. 7, $f_t(T)$ is equivalent to the distribution of recovery timescales. Note that for $c = 0.5$ the distribution shifts rightward as a function of time, as explained following Eq. 36.

cific channel model with power-law RTPDF, recall that the voltage input influences the channel through the variables $c$ and $\gamma$, which are functions of the voltage.

### Generalization of the time-dependent rate

One way to derive the general dynamic equation, is to generalize the definition of the time-dependent rate so that it also depends on the 'global time' $t$,

$$\lambda(t \mid u) \triangleq \lim_{\Delta t \to 0} \frac{1}{\Delta t} P(t_s \leq t + \Delta t \mid t_s \geq t \geq u),$$

where $t$ is the current time, $u$ is the time of the last transition between states, and $t_s$ is the time of the next switch in channel states. The probability that the channel will switch back to the other state in the time interval $(t, t')$ can be calculated by dividing $(t, t')$ into $n$ smaller time segments $(t_i, t_i + \Delta t), i = 1, 2, \ldots, n$. Breaking up the interval $(t, t')$ into these segments, we have,

$$P(t_s \leq t' \mid t_s \geq t \geq u)$$

$$= \lim_{\Delta t \to 0} \sum_{i=1}^{n} \left( \prod_{k=1}^{i-1} \left( 1 - \lambda(t_k \mid u)\Delta t + O\left(\left(\lambda(t_k \mid u)\Delta t\right)^2\right) \right) \right) \lambda(t_i \mid u)\Delta t$$

$$= \lim_{\Delta t \to 0} \sum_{i=1}^{n} \exp\left( \sum_{k=1}^{i-1} \ln\left( 1 - \lambda(t_k \mid u)\Delta t + O\left(\left(\lambda(t_k \mid u)\Delta t\right)^2\right) \right) \right) \lambda(t_i \mid u)\Delta t$$

$$= \lim_{\Delta t \to 0} \sum_{i=1}^{n} \exp\left( -\sum_{k=1}^{i-1} \left( \lambda(t_k \mid u)\Delta t + O\left(\left(\lambda(t_k \mid u)\Delta t\right)^2\right) \right) \right) \lambda(t_i \mid u)\Delta t$$

$$= \int_{t}^{t'} \exp\left( -\int_{t}^{v} \lambda(z \mid u)dz \right) \lambda(v \mid u)dv$$

$$= 1 - \exp\left( -\int_{t}^{t'} \lambda(z \mid u)dz \right),$$

where we have defined $\Pi_{k=1}^{0} = 1$, $\Sigma_{k=1}^{0} = 0$. The last equality can easily be established by taking the derivative of the final term with respect to $t'$.

We have thus shown that:

$$P(t_s \leq t' \mid t_s \geq t \geq u) = 1 - \exp\left( -\int_{t}^{t'} \lambda(z \mid u)dz \right) \quad (41)$$

Next we introduce a 'time-dependent RTPDF', which can be thought of as an extension of the RTPDF of the type introduced in Section 'The Model'. This function is the probability density to switch back into the other state at time $x$:

$$\psi(x \mid y) \triangleq \frac{d}{dx} \left[ P(t_s \geq x \mid t_s \geq t = u = y) \right]$$

$$= \exp\left( -\int_{y}^{x} \lambda(z \mid y)dz \right) \lambda(x \mid y). \quad (42)$$

Now that we have defined the time-dependent RTPDFs in Eq. 42, we can generalize the dynamic equation developed in Eq. 16 (through the projection method). Setting $p(0) = 1$ for simplicity, we get,

$$\frac{d}{dt} p(t) = -\gamma(t) p(t) + \int_{0}^{t} \gamma(u) p(u) \psi_t(t \mid u) du, \quad (43)$$

which allows us to calculate the behavior of $p(t)$ for any time-varying input. Notice that the term $\int_{0}^{\infty} \pi_\delta(0) \psi_t(t) d\delta$ that appeared in Eq. 16, has been removed because we chose $p(0) = 1$. Since the input voltage is now general, choosing $p(0) = 1$ is less of a constraint than before, as we can simply choose $t = 0$ to be some time at which the channel was available. By doing so we got rid of the initial conditions defined by $\pi_\delta(0)$, which are unobservable.

### The case of power-law RTPDF

For a constant rate $\lambda(t|u) = \gamma$ we get, as expected, the Markovian exponential RTPDF,

$$\psi_E(t\,|\,u) = \gamma \exp\left(-\gamma(t-u)\right) \quad (t > u),$$

while for a varying rate $\gamma(t)$ we get,

$$\psi_E(t\,|\,u) = \gamma(t) \exp\left(-\int_u^t \gamma(z)\,dz\right) \quad (t > u). \tag{44}$$

For the the non-Markovian state with a constant input $\lambda(t|u) = c/(t - u + t_0)$, we get as expected,

$$\psi_P(t\,|\,u) = \exp\left(-\int_u^t \frac{c}{z-u+t_0}\,dz\right) \frac{c}{t-u+t_0}$$
$$= \frac{c/t_0}{(1+(t-u)/t_0)^{c+1}} \quad (t > u),$$

while for general input, if we assume that

$$\lambda(t\,|\,u) = \frac{c(t)}{t-u+t_0}, \tag{45}$$

we obtain

$$\psi_P(t\,|\,u) = \exp\left(-\int_u^t \frac{c(z)}{z-u+t_0}\,dz\right) \frac{c(t)}{t-u+t_0} \quad (t > u), \tag{46}$$

For example, if $c(t) = c_1$ for $t < t_1$ and $c(t) = c_2$ for $t \geq t_1$, then we can write, for all $t \geq t_1$,

$$\psi_P(t\,|\,u) = \exp\left(-\int_u^t \frac{c(z)}{z-u+t_0}\,dz\right) \frac{c(t)}{t-u+t_0}$$
$$= \begin{cases} \dfrac{c_2}{t_0} \dfrac{\left(1+(t-u)/t_0\right)^{c_2-c_1}}{\left(1+(t-u)/t_0\right)^{c_2+1}}, & u < t_1 \\[2ex] \dfrac{c_2}{t_0} \dfrac{1}{\left(1+(t-u)/t_0\right)^{c_2+1}}, & u \geq t_1 \end{cases},$$

and so we get the following dynamic equation,

$$\frac{d}{dt} p(t) = -\gamma(t)p(t) + \int_{t_1}^t \gamma(u)p(u)\psi_P(t-u)\,du$$
$$+ \int_0^{t_1} \gamma(u)p(u)\left(1+(t_1-u)/t_0\right)^{c_2-c_1}\psi_P(t-u)\,du, \tag{47}$$

where we denoted $\psi_P(t) = (c_2/t_0)(1+t/t_0)^{-c_2-1}$.

And so, the initial conditions for the evolution of $p(t)$ from time $t_1$ onward, are completely determined by the last term in Eq. 47. Also, it is important to note that the expression on Eq. 47 is easier to calculate numerically than Eq. 43, since all the integrals can be written in a convolution form.

### Direct generalization of the RTPDF

Another way to generalize the channel model to include varying input is to directly generalize the definition of the RTPDF, instead of going through the time-dependent rate. This could be done if we assume that the two-state non-Markov model is based on some underlying many-states Markov model, as explained in Section 'How to create a two-state non-Markov model from a general Markov model'. In the case of time-varying rates, it is straightforward to generalize Eqs 30 and 31 and get

$$\psi_A(t\,|\,u) = \gamma(t) \exp\left(-\int_u^t \gamma(z)\,dz\right), \quad t \geq 0, \tag{48}$$

$$\psi_I(t\,|\,u) = \bar{\pi}_A^s(u)\mathbf{G}_{AI}(u) \exp\left(\int_u^t \mathbf{Q}_{II}(z)\,dz\right)\mathbf{Q}_{IA}(t)\bar{u}_A^T, \quad t \geq 0, \tag{49}$$

where we defined $\gamma(t) = \bar{\pi}_A^s(t)\mathbf{Q}_{AI}(t)\bar{u}_I^T$, $(\mathbf{G}_{AI}(t))_{mn} = (\mathbf{Q}_{AI}(t))_{mn} / (\sum_{l=1}^L (\mathbf{Q}_{AI}(t))_{ml})$ and $\bar{\pi}_A^s(t)$ is the stationary distribution in the available set of states at time $t$: when all the transitions to inactivation are forbidden, and we assume that the input changes slower than time it takes to reach the stationary distribution over the available states. We can substitute Eq. 49 into Eq. 43 to get the dynamic equation for $p(t)$, and derive an expression for the generalized time-dependent rate (similarly to the derivation in Eqs 32 and 33), yielding

$$\lambda_I(t\,|\,u) = \frac{\psi_I(t\,|\,u)}{\int_t^\infty \psi_I(z\,|\,u)\,dz} = \frac{\psi_I(t\,|\,u)}{1 - \int_u^t \psi_I(z\,|\,u)\,dz} \tag{50}$$

where the rightmost expression allows to calculate $\lambda_I(t|u)$ in a causal way, without knowing future inputs. Notice that here $\lambda_I(t|u)$ may generally depend on the entire history of the input since time $u$.

For example, suppose we use this method on the same state space described in the case of projection method (see 'Projection' of the Inactivated State) and assume that only the rates of inactivation change with voltage. Taking the continuum limit of Eq. 49, it is straightforward to get the following generalization of Eq. 12, in the power-law case:

$$\psi_P(t\,|\,u) = \int_0^\infty f_u(\delta)\delta \exp(-\delta(t-u))\,d\delta$$
$$= \frac{c(u)/t_0}{(1+(t-u)/t_0)^{c(u)+1}} \tag{51}$$

where $f_u(\delta) = (t_0^{c(u)}/\Gamma(c(u)))\delta^{c(u)-1}\exp(-\delta t)$, $\delta > 0$ is obvious generalization of Eq. 18. Using Eq. 50 we get,

$$\lambda_P(t\,|\,u) = \frac{c(u)}{t-u+t_0} \tag{52}$$

Notice that Eq. 52 differs from Eq. 45, and also that Eq. 51 differs from Eq. 46. Recall that Eq. 45. was an assumption, which we used to derive the RTPDF in 46. In contrast, Eq. 52 was derived from the RTPDF at Eq. 51, which was itself derived from the assumed Markovian model of the projection method. The different assumptions led to different results. Specifically, this shows that when we take into account time-varying inputs, the Markov model used in the projection method is no longer equivalent to all other models. In this work we chose to model the channel behavior through Eq. 45, which is easier to implement numerically. Using Eq. 52 instead should not significantly change the results for the various inputs examined in this work (constant input, step input, oscillating input and slowly changing input) though it should have an effect for other types of inputs. The final choice between Eqs 45 and 52 or some other, more complicated expression, should be made through carefully designed experiments on channels.

## OSCILLATING VOLTAGE INPUT

We examine the case where the input to the model oscillates with period $T_p$. These oscillations may arise, for example, from periodic stimulation or some noise process (in which case $T_p$ is only approximate). For example, in the context of the specific channel model with power-law RTPDF the parameters $\gamma$, $c$ may generally be voltage-dependent and vary with time, while $t_0$ is maintained constant. Intuitively, if $T_p$ is sufficiently small in comparison with the transition timescales, then on sufficiently large timescales we should not notice these fluctuations in the parameters (even though they may be large in magnitude), and expect them to 'average-out'. Here we prove this claim rigorously, and find the exact conditions for this to happen.

For any time-dependent variable $X_s$ define the mean over a period $T_p$,

$$\hat{X}_t \triangleq \frac{1}{T_p} \int_t^{t+T_p} X_z \, dz.$$

(53)

Again, $t$ denotes the current time, $u$ is the time of the last transition between states, and $t_s$ is the time of the next switch in channel states. Using Eq. 41, we can write, for all $\theta > 0$,

$$P\left(t_S \leq t + \theta \,|\, t_S \geq t \geq u\right)$$
$$= 1 - \exp\left(-\int_t^{t+\theta} \lambda(z \,|\, u) dz\right)$$
$$= \int_t^{t+\theta} \lambda(z \,|\, u) dz + O\left(\left(\int_t^{t+\theta} \lambda(z \,|\, u) dz\right)^2\right)$$
$$= \hat{\lambda}(t \,|\, u)\theta + O\left(\int_t^{t+\theta} \left(\lambda(z \,|\, u) - \hat{\lambda}(t \,|\, u)\right) dz\right) + O\left(\left(\hat{\lambda}(t \,|\, u)\theta\right)^2\right),$$

where, in order to obtain the final line, we replaced $\lambda(z|u)$ by $\hat{\lambda}(t \,|\, u) + (\lambda(z \,|\, u) - \hat{\lambda}(t \,|\, u))$ in the penultimate line. From the above derivation, if we can choose $\theta$ so that $\hat{\lambda}(t \,|\, u)\theta \ll 1$, and $\int_t^{t+\theta} |\lambda(z \,|\, u) - \hat{\lambda}(t \,|\, u)| du \ll \hat{\lambda}(t \,|\, u)\theta$, we can safely approximate,

$$P\left(t_S \leq t + \theta \,|\, t_S \geq t \geq u\right) \approx \hat{\lambda}(t \,|\, u)\theta.$$

(54)

Then, for every timescale above $\theta$ we can replace $\lambda(t|u)$ by $\hat{\lambda}(t \,|\, u)$.

### The case of power-law RTPDF

In the context of our model, this means that if, for all $t$, $T_p \ll \gamma^{-1}(t)$, we can approximate,

$$P\left(t_I \leq t + \theta \,|\, t_I \geq t \geq u\right) \approx \dot{\hat{\gamma}}\theta,$$

(55)

where $t_I$ is the time of inactivation, and the index $t$ was suppressed in $\dot{\hat{\gamma}}$, since it is assumed to be, at least approximately, constant.

Also, if $T_p \ll \min_s [t_0/c(s)]$ and $T_p \ll t_0$ then we can approximate, using Eq. 45,

$$P\left(t_R \leq t + \theta \,|\, t_R \geq t \geq u\right) \approx \frac{\hat{c}}{t + \theta - u + t_0}\theta,$$

(56)

where again, $t_R$ is the time of recovery, and the $t$ index was suppressed in $\hat{c}$.

In conclusion, for rapidly oscillating input, we can replace $\gamma$, $c$ by their time-averaged expressions $\dot{\hat{\gamma}}, \hat{c}$.

## THE JOINT PROBABILITY DISTRIBUTION OF THE AVAILABILITY

So far we have dealt only with the marginal probability to be in the available state – $p(t)$. To complete the description of the process, we discuss here joint probability distributions.

The available state is Markovian, which makes it easy to derive $p(t_1, t_2, \ldots, t_k)$, the joint probability distribution to be in the available state at some arbitrary times $t_1, \ldots, t_k$,

$$p(t_1, t_2, \ldots, t_k) = p(t_1) p(t_2 \,|\, t_1) \cdots p(t_k \,|\, t_{k-1}),$$

where we denoted $p(t_i|t_j)$ as the probability of the channel to be in the available state at time $t_i$ assuming that at $t_j$ the channel was also in the available state. For arbitrary voltage input, each $p(t_i|t_j)$ term in this product is the solution of Eq. 43, with the initial condition $p(t_j) = 1$ (notice also that the lower limit of integration in Eq. 43 must be set to $t_j$). In the case that the voltage is constant, $p(t_i|t_j) = p(t_i - t_j)$, where $p(t)$ is the solution of Eq. 16 with the initial condition $p(0) = 1$. If we assume also, for simplicity, that $p(0) = 1$, we get,

$$p(t_1, t_2, \ldots, t_k) = p(t_1) p(t_2 - t_1) \cdots p(t_k - t_{k-1})$$

where $p(t)$ is the solution of Eq. 16.

Next, we compute the joint moments and the auto-covariance function. Define $S(t)$ to be the channel state at time $t$ and introduce the indicator function $\mathcal{I}(t)$ to equal 1 if, and only if, $S(t) = A$, and zero otherwise, then,

$$\langle \mathcal{I}(t_1) \mathcal{I}(t_2) \cdots \mathcal{I}(t_k) \rangle = p(t_1, t_2, \ldots, t_k),$$

where the angular brackets indicate an average. Thus the joint distributions are equal to the joint moments of the available state.

Using the above results, we can easily calculate $k(t_1, t_2)$, the availability auto-covariance function, where we assume that $t_2 > t_1$,

$$k(t_1, t_2) \triangleq \langle \mathcal{I}(t_1) \mathcal{I}(t_2) \rangle - \langle \mathcal{I}(t_1) \rangle \langle \mathcal{I}(t_2) \rangle,$$
$$= p(t_1) \left( p(t_2 \,|\, t_1) - p(t_2) \right).$$

### The case of power-law RTPDF

For example, in the context of our channel model, and for constant voltage input, we use Eq. 4 and asymptotically obtain, for $c > 1$,

$$k(t) = p_\infty^2 \left(1 - p_\infty\right) t_0^{c-1} \cdot t^{-|1-c|} \tag{57}$$

where $k(t_2 - t_1) = k(t_1, t_2)$ is the stationary auto-covariance.

## ACKNOWLEDGMENTS

The authors are grateful to Erez Braun, Naama Brenner, Asaf Gal and Avner Wallach for many insightful discussions. Special thanks to Yuval Elhanati, Yariv Kafri and Shimon Marom for the considerable effort they invested in improving the quality and focus of this manuscript. The work of R. Meir is partially supported by grant No. 665/08 from the Israel Science Foundation.

## SUPPLEMENTARY MATERIAL

**Conflict of Interest Statement:** The authors declare that the research was conducted in the absence of any commercial or financial relationships that could be construed as a potential conflict of interest.

*Received: 14 December 2009; paper pending published: 04 February 2010; accepted: 02 March 2010; published online: 08 April 2010.*
*Citation: Soudry D and Meir R (2010) History-dependent dynamics in a generic model of ion channels – an analytic study. Front. Comput. Neurosci. 4:3. doi: 10.3389/fncom.2010.00003*

# History-dependent dynamics in a generic model of ion channels – an analytic study

## Daniel Soudry[1,2] and Ron Meir[1,2]*

[1] Department of Electrical Engineering, Technion, Haifa, Israel
[2] Laboratory for Network Biology Research, Technion, Haifa, Israel

Edited by:     David Hansel, University of Paris, France

Reviewed by:   Nicolas Brunel, Centre National de la Recherche Scientifique, France
Carl van Vreeswijk, Centre National de la Recherche Scientifique, France

*Correspondence: Ron Meir, Department of Electrical Engineering, Technion, Haifa 32000, Israel. e-mail: rmeir@ee.technion.ac.il

## THE RELATION BETWEEN $p(t)$ AND $A(t)$

In Section 'Background' we claim that $A(t)$, the total availability of a channel population under voltage clamp, is approximately equal to $p(t)$, the probability of a single channel to be available. Here we show why this claim is true. Also, we explain the relation between the auto-covariance function of a single channel and a population of channels, which was mentioned in Section 'Temporal Correlations'.

We define $S_n(t)$ to be the state of the $n$-th channel, and the function:

$$\mathcal{I}_n(t) = \begin{cases} 1 & \text{if} \quad S_n(t) = A \\ 0 & \text{if} \quad S_n(t) = I \end{cases},$$

which indicates the availability of the $n$-th channel.

The total availability at time $t$, which is the fraction of the available channels, is

$$A(t) = \frac{1}{N} \sum_{n=1}^{N} \mathcal{I}_n(t),$$

where $N$ is the size of the population.

From the law of the large numbers, when the channel population is very large, we get:

$$\lim_{N \to \infty} A(t) = \lim_{N \to \infty} \frac{1}{N} \sum_{n=1}^{N} \mathcal{I}_n(t) = \langle \mathcal{I}_1(t) \rangle = p(t),$$

where we use angular brackets to denote averages. For any population size $N$, we get:

$$\langle A(t) \rangle = \frac{1}{N} \sum_{n=1}^{N} \langle \mathcal{I}_n(t) \rangle = \frac{1}{N} \sum_{n=1}^{N} p(t) = p(t).$$

Defining $k_N(t_1, t_2)$ he the finite sample auto-covariance function of the channel population, we have:

$$\begin{aligned}
k_N(t_1, t_2) \\
= \langle (A(t_1) - p(t_1))(A(t_2) - p(t_2)) \rangle \\
= \left\langle \left( \frac{1}{N} \sum_{n=1}^{N} (\mathcal{I}_n(t_1) - p(t_1)) \right) \left( \frac{1}{N} \sum_{n=1}^{N} (\mathcal{I}_n(t_2) - p(t_2)) \right) \right\rangle \\
= \frac{1}{N^2} \sum_{n=1}^{N} \langle (\mathcal{I}_n(t_1) - p(t_1))(\mathcal{I}_n(t_2) - p(t_2)) \rangle \\
+ \frac{1}{N^2} \sum_{n=1}^{N} \sum_{\substack{m=1 \\ m \neq n}}^{N} \langle (\mathcal{I}_n(t_1) - p(t_1)) \rangle \langle (\mathcal{I}_m(t_2) - p(t_2)) \rangle \\
= \frac{1}{N^2} \sum_{n=1}^{N} k(t_1, t_2) + \frac{1}{N^2} \sum_{n=1}^{N} \sum_{\substack{m=1 \\ m \neq n}}^{N} 0 \\
= \frac{1}{N} k(t_1, t_2),
\end{aligned}$$

where $k(t_1, t_2)$ is the single channel auto-covariance function, a and we used the fact the different channels are independent, under voltage clamp.

Specifically,

$$\text{Var}(A(t)) = k_N(t, t) = \frac{1}{N} k(t, t) = \frac{1}{N} \text{Var}(S_1(t)) = \frac{1}{N} p(t)(1 - p(t)).$$

and thus, the coefficient of variation is

$$CV_{A(t)} = \frac{\sigma_{A(t)}}{\langle A(t) \rangle} = \frac{\sqrt{\text{Var}(A(t))}}{p(t)} = \frac{\sqrt{(1/p(t) - 1)}}{N}$$

which goes to zero as $N \to \infty$, as expected from the law of large numbers.

## THE MARKOVIAN LIMIT OF THE NON-MARKOVIAN STATE

In Section 'Model Description' we defined the RTPDF of the non-Markovian state to be:

$$\psi_P(t) = \frac{c/t_0}{(1 + t/t_0)^{c+1}}$$

If we naively take the limit that $c \to \infty$, then $\psi_p(t) \to \delta(t)$, a Dirac delta function at zero, meaning that the channel will recover instantly. And so, in order for this limit to be meaningful, we need to 'zoom in' in the time axis to see what happens for very short times. To do this, we change the variables of the RTPDF $-\theta \triangleq c \cdot t$, so that:

$$\overline{\psi}_P(\theta) = \psi_P(t) \cdot \frac{dt}{d\theta} = \psi_P(\theta/c)\frac{1}{c}.$$

This gives us:

$$\lim_{c \to \infty} \overline{\psi}_P(\theta) = \lim_{c \to \infty} \frac{1/t_0}{(1 + \theta/(t_0c))^{c+1}},$$

$$= \lim_{c \to \infty} \frac{1/t_0}{\left[(1 + \theta/(t_0c))^{t_0c/\theta}\right]^{\theta/t_0}},$$

$$= \frac{1}{t_0} \exp(-\theta/t_0),$$

where we used the limit $\lim_{x \to \infty}(1 + 1/x)^x = e$ in the last line.

In conclusion: in the limit of $c \to \infty$ the non-Markovian state approximates a Markovian state, for very short times.

## CALCULATION OF $\tilde{\psi}_P(s)$

In this section we develop an expression for $\tilde{\psi}_P(s)$, the Laplace transform of $\psi_P(t)$. This will give Eq. 19, which is the basis for the derivations in Section 'Asymptotic Solution of the Dynamic Equations for Power-Law RTPDF'. By the definition of the Laplace transform:

$$\tilde{\psi}_P(s) = \frac{c}{t_0} \int_0^\infty \frac{e^{-st}dt}{(1 + t/t_0)^{c+1}}$$

$$\overset{1}{=} ct_0^c \int_0^\infty \frac{e^{-re^{i\omega}(t+t_0)}dt}{(t+t_0)^{c+1}}$$

$$\overset{2}{=} ct_0^c e^{st_0} r^c \int_{st_0}^\infty \frac{e^{-e^{i\omega}z}dz}{z^{c+1}}$$

$$= ct_0^c e^{st_0} r^c \int_{st_0}^\infty \left[\sum_{k=0}^\infty \frac{z^{k-c-1}}{k!}\left(-e^{i\omega}\right)^k\right] dz$$

$$\overset{3}{=} ct_0^c e^{st_0} r^c \left[F(\omega) - \sum_{k=0}^\infty \frac{(rt_0)^{k-c}}{k!(k-c)}\left(e^{i\omega}\right)^k\right]$$

$$= ce^{st_0}\left[F(\omega)(t_0r)^c - \sum_{k=0}^\infty \frac{(-st_0)^k}{k!(k-c)}\right]$$

$$\overset{4}{=} ce^{st_0}\left[G \cdot (t_0s)^c - \sum_{k=0}^\infty \frac{(-st_0)^k}{k!(k-c)}\right] \tag{S1}$$

The numbered steps are explained as follows:

1. This transform is properly defined for all $s$ such as $\mathrm{Re}[s] \geq 0$. And so, we denoted $s \equiv re^{i\omega}$ for some $r \geq 0, \omega \varepsilon(-\frac{\pi}{2}, \frac{\pi}{2})$.
2. We denoted $z \equiv r(t + t_0)$.
3. We assumed here that $c$ is not an integer, used Fubini's theorem (which we can use since $\int_{st_0}^\infty |e^{-e^{i\omega}z}dz/z^{c+1}| < \infty$) to switch the order of summation and integration, and denoted $F(\omega)$ to be some function of $\omega$.
4. We used the fact that $\tilde{\psi}(s)$ must be piecewise analytic since for every $n$:

$$\frac{d^n}{ds^n}\tilde{\psi}_P(s) = \mathcal{L}_+\left[(-t)^n \frac{c/t_0}{(1 + t/t_0)^{c+1}}\right]$$

which always exists $\forall \mathrm{Re}[s] > 0$ — therefore $F(\omega)$ $(t_0r)^c$ must be a function of $s$ — and the only one possible is $G$ $(t_0s)^c$, where $G$ is some constant.

If we assume in step 3 that $c$ is an integer, than we get instead:

$$\tilde{\psi}_P(s) = ce^{st_0}\left[G \cdot (t_0s)^c - \frac{1}{c!}(-st_0)^c \cdot \ln(st_0) - \sum_{\substack{k=0 \\ k \neq c}}^\infty \frac{(-st_0)^k}{k!(k-c)} \cdot\right] \tag{S2}$$

Also, since we can write:

$$\tilde{\psi}_P(s) = ct_0^c s^c e^{st_0} \int_{st_0}^\infty \frac{e^{-z}dz}{z^{c+1}}$$

$$= ct_0^c s^c e^{st_0} \Gamma(-c, st_0)$$

we can compare this derivation (which was done for complex $s$) with the one in Bender and Orszag (1978: 251–252) where it is shown that $G = \Gamma(-c) < 0$, in the case that $c$ is not an integer, and if $c$ is an integer then:

$$G = -\frac{(-1)^{N+1}}{N!}\left(\gamma - \sum_{n=1}^N \frac{1}{n}\right)$$

where $\gamma$ is Euler constant (do not confuse it with the inactivation rate). From the analyticity of $\tilde{\psi}_P(s) \forall \mathrm{Re}[s] > 0$ we come to the conclusion that $G$ is identical also in our case.

## ASYMPTOTIC BEHAVIOR OF $\tilde{\psi}_P(s)$

In this section we examine the expressions for $\tilde{\psi}_P(s)$ which were derived in Eqs S1 and S2 for various values of $c$ and retain the lowest orders in $s$. For non-integer values of $c$ this will give Eq. 20, which appears in Section 'Asymptotic Solution of the Dynamic Equations for Power-Law RTPDF'.

When $c$ is not an integer:

$$\tilde{\psi}_P(s) = ce^{st_0}\left[G \cdot (t_0s)^c - \sum_{k=0}^\infty \frac{(-st_0)^k}{k!(k-c)}\right]$$

$$= c\left(1 + st_0\sigma_{A(t)} + \frac{1}{2}s^2t_0^2 + O(s^3)\right)$$

$$\times \left(G \cdot (t_0s)^c + \frac{1}{c} + \frac{1}{1-c}st_0 - \frac{1}{2(2-c)}s^2t_0^2 + O(s^3)\right)$$

$$= 1 + \frac{c}{1-c} st_0 - \frac{c}{2(2-c)} s^2 t_0^2 + cG \cdot (t_0 s)^c + cG \cdot (t_0 s)^{c+1}$$
$$+ st_0 + \frac{c}{1-c} s^2 t_0^2 + \frac{1}{2} s^2 t_0^2 + O(s^3)$$
$$= 1 + \frac{1}{1-c} st_0 + \frac{1}{(c-2)(c-1)} s^2 t_0^2 + cG \cdot (t_0 s)^c + O\left(s^{\min(3,c+1)}\right)$$

For $c = 1$,

$$\tilde{\psi}_P(s) = e^{st_0}\left[ G \cdot st_0 + st_0 \cdot \ln(st_0) - \sum_{\substack{k=0 \\ k \neq 1}}^{\infty} \frac{(-st_0)^k}{k!(k-1)} \right]$$
$$= \left(1 + st_0 + \frac{1}{2} s^2 t_0^2 + O(s^3)\right)$$
$$\times \left(1 + st_0 \cdot \ln(st_0) + G \cdot st_0 - \frac{1}{2} s^2 t_0^2 + O(s^3)\right)$$
$$= 1 + st_0 \cdot \ln(st_0) + (G+1) \cdot st_0 + (st_0)^2 \cdot \ln(st_0)$$
$$+ (G \cdot st_0)^2 + O(s^3 \ln(st_0)).$$

For $c = 2$,

$$\tilde{\psi}_P(s) = 2e^{st_0}\left[ G \cdot (st_0)^2 - \frac{1}{2}(st_0)^2 \cdot \ln(st_0) - \sum_{\substack{k=0 \\ k \neq 2}}^{\infty} \frac{(-st_0)^k}{k!(k-2)} \right]$$
$$= 2\left(1 + st_0 + \frac{1}{2} s^2 t_0^2 + O(s^3)\right)$$
$$\times \left(\frac{1}{2} - st_0 - \frac{1}{2}(st_0)^2 \cdot \ln(st_0) + G \cdot (st_0)^2 + O(s^3)\right)$$
$$= 1 - st_0 - (st_0)^2 \cdot \ln(st_0) + \left(2G - \frac{3}{2}\right) \cdot (st_0)^2 + O(s^3 \ln(st_0)).$$

For $c = 3, 4, 5, \ldots$

$$\tilde{\psi}_P(s) = ce^{st_0}\left[ G \cdot (st_0)^c - \frac{1}{c!}(st_0)^c \cdot \ln(st_0) - \sum_{\substack{k=0 \\ k \neq c}}^{\infty} \frac{(-st_0)^k}{k!(k-c)} \right]$$
$$= c\left(1 + st_0 + \frac{1}{2} s^2 t_0^2 + O(s^3)\right)$$
$$\times \left(\frac{1}{c} + \frac{1}{1-c} st_0 - \frac{1}{2(2-c)} s^2 t_0^2 - \frac{1}{c!}(st_0)^c \cdot \ln(st_0) + O(s^3)\right)$$
$$= 1 + \frac{1}{1-c} st_0 + \frac{1}{(c-2)(c-1)} s^2 t_0^2 - \frac{1}{c!}(st_0)^c \cdot \ln(st_0) + O(s^3).$$

## ASYMPTOTIC BEHAVIOR OF $\tilde{p}(s)$ FOR INTEGER VALUES OF $c$

In Section 'Asymptotic Solution of the Dynamic Equations for Power-Law RTPDF' we derive an asymptotic solution for $\tilde{p}(s)$ in the case that $c$ is not an integer. In this section we show a similar derivation for the integer case.

Assuming $p(0) = 1$, we use the Eq. 17:

$$\tilde{p}(s) = \frac{1}{s + \gamma\left(1 - \tilde{\psi}_P(s)\right)},$$

and substitute $\tilde{\psi}_P(s)$ in the expression for the case $c = 1$,

$$\tilde{p}(s) = \frac{1}{\gamma st_0 \cdot \ln(st_0) + O(s)}$$
$$= \frac{1}{\gamma st_0 \cdot \ln(st_0)} + O\left(\frac{1}{s \cdot \ln^2(st_0)}\right).$$

For $c = 2$ we get:

$$\tilde{p}(s) = \frac{1}{(1+\gamma)st_0 + (st_0)^2 \cdot \ln(st_0) - (2G - \frac{3}{2}) \cdot (st_0)^2 + O(s^3 \ln(st_0))}$$
$$= \frac{1}{(1+\gamma)st_0} \frac{1}{1 + \frac{1}{1+\gamma} st_0 \ln(st_0) - \frac{1}{1+\gamma}(2G - \frac{3}{2}) \cdot st_0 + O(s^2 \ln(st_0))}$$
$$= \frac{1}{(1+\gamma)st_0} + \frac{1}{(1+\gamma)^2} \ln(st_0) + O(1)$$

For $c = 3, 4, 5, \ldots$ we get:

$$\tilde{p}(s) = \frac{1}{s - \frac{\gamma}{1-c} st_0 - \frac{\gamma}{(c-2)(c-1)} s^2 t_0^2 + \frac{\gamma}{c!}(st_0)^c \cdot \ln(st_0) + O(s^3)}$$
$$= p_\infty s^{-1} \cdot \frac{1}{1 - p_\infty \frac{\gamma}{(c-2)(c-1)} st_0^2 + p_\infty \frac{\gamma}{c!}(st_0)^{c-1} \cdot \ln(st_0) + O(s^3)}$$
$$= p_\infty s^{-1} + p_\infty^2 \frac{\gamma}{c!}(st_0)^{c-2} \cdot \ln(st_0) + O(1)$$

where,

$$p_\infty = \begin{cases} \frac{c-1}{\gamma t_0 + c - 1} & \text{if } c \geq 1, \\ 0 & \text{if } 0 < c < 1, \end{cases}$$

is, as in the non-integer case, the steady state of $p(t)$ obtained from the from the final value theorem. Unfortunately, since the Tauberian theorem does not apply to logarithmic expressions we cannot apply it to infer the asymptotic behavior of $p(t)$. From numerical simulations we see that the present case (integer valued $c$) is not pathological, and we can simply take it as the limit of the non-integer case.

## THE MOMENTS OF $f_t(T)$

In this section we derive Eqs 37–40.

To do this we need calculate the moments of Eq. 36:

$$f_t(T) = \begin{cases} \frac{c-1}{t_0}(1 + T/t_0)^{-c} & \text{if } 1 < c \\ \frac{\sin(\pi c)}{\pi t_0^c}(t - T)^{c-1} & (1 + T/t_0)^{-c} & \text{if } 0 < c < 1 \end{cases}, (0 \leq T \leq t) \quad (S3)$$

First we examine the case $1 < c$

The mean:

$$\langle T \rangle = \int_{-\infty}^{\infty} T f_r(T) dT$$

$$= \frac{c-1}{t_0} \int_0^t \frac{T}{(1+T/t_0)^c} dT$$

$$= \frac{1}{t_0(c-2)} \left[ t_0^2 - \left(\frac{1}{1+t/t_0}\right)^c (t+t_0)(t_0+(c-1)t) \right]$$

The behavior of the last expression in the limit $t \to \infty$ is determined by the value of $c$.

$$\langle T \rangle = \begin{cases} \frac{c-1}{2-c} t_0^{c-1} \cdot t^{2-c} + O(1) & \text{if } 1 < c < 2 \\ \frac{t_0}{c-2} + O(t^{2-c}) & \text{if } c > 2 \end{cases}$$

The second moment:

$$\langle T^2 \rangle = \int_{-\infty}^{\infty} T^2 f_r(T) dT$$

$$= \frac{c-1}{t_0} \int_0^t \frac{T^2}{(1+T/t_0)^c} dT$$

$$= \frac{1}{t_0(c-2)(c-3)} \left[ 2t_0^3 - \left(\frac{1}{1+t/t_0}\right)^c (t+t_0)(2t_0^2 \right.$$

$$\left. + 2(c-1)t_0 t + (c-1)(c-2)t^2) \right]$$

The behavior of the final expression in the limit $t \to \infty$ is again determined by the value of $c$.

$$\langle T^2 \rangle = \begin{cases} \frac{c-1}{3-c} t_0^{c-1} \cdot t^{3-c} + O(1) & \text{if } 1 < c < 3 \\ \frac{2t_0^2}{(c-2)(c-3)} + O(t^{3-c}) & \text{if } c > 3 \end{cases}$$

so we get:

$$\sigma_T = \sqrt{\langle T^2 \rangle \langle T \rangle^2}$$

$$= \begin{cases} \sqrt{\frac{c-1}{3-c} t_0^{c-1} \cdot t^{3-c} - \left(\frac{c-1}{2-c}\right)^2 t_0^{2c-2} \cdot t^{4-2c} + O(1)}, & 1 < c < 2 \\ \sqrt{\frac{c-1}{3-c} t_0^{c-1} \cdot t^{3-c} - \left(\frac{t_0}{c-2}\right)^2 + O(1)}, & 2 < c < 3 \\ \sqrt{\frac{2t_0^2}{(c-2)(c-3)} - \left(\frac{t_0}{c-2}\right)^2 + O(t^{3-c})}, & c > 3 \end{cases}$$

$$= \begin{cases} \sqrt{\frac{c-1}{3-c}} t_0^{(c-1)/2} \cdot t^{(3-c)/2} + O(t^{2-c}), & 1 < c < 2 \\ \sqrt{\frac{c-1}{3-c}} t_0^{(c-1)/2} \cdot t^{(3-c)/2} + O(1), & 2 < c < 3 \\ \frac{t_0}{c-2} \sqrt{\frac{c-1}{c-3}} + O(t^{(3-c)/2}), & c > 3 \end{cases}$$

and,

$$\frac{\sigma_T}{\langle T \rangle} = \begin{cases} \frac{2-c}{\sqrt{(3-c)(c-1)}} t_0^{(1-c)/2} \cdot t^{(c-1)/2} + O(1), & 1 < c < 2 \\ (c-2) \sqrt{\frac{c-1}{3-c}} t_0^{(c-3)/2} \cdot t^{(3-c)/2} + O(1), & 2 < c < 3 \\ \sqrt{\frac{c-1}{c-3}} + O(t^{(3-c)/2}), & c > 3 \end{cases}$$

Next, we handle the case $0 < c < 1$.

$$\langle T^k \rangle = \int_{-\infty}^{\infty} T^k f_r(T) dT$$

$$= \frac{\sin(\pi c)}{\pi t_0^c} \int_0^t \frac{1}{(t-T)^{1-c}} \cdot \frac{T^k}{(1+T/t_0)^c} dT$$

$$= \frac{\sin(\pi c)}{\pi t_0^c} t^{c-1} \int_0^t \frac{1}{(1-T/t)^{1-c}} \cdot \frac{T^k}{(1+T/t_0)^c} dT$$

$$= \frac{\sin(\pi c)}{\pi t_0^c} t^{c+k} \int_0^1 \frac{1}{(1-x)^{1-c}} \cdot \frac{x^k}{(1+tx/t_0)^c} dx$$

$$= \frac{\sin(\pi c)}{\pi} t^k \int_0^1 (1-x)^{c-1} \cdot x^{k-c} dx + O(t^{k-1})$$

$$= \frac{\sin(\pi c)}{\pi} t^k \cdot B(k-c+1, c) + O(t^{k-1})$$

$$= \frac{-1}{c\Gamma(-c)\Gamma(c)} t^k \cdot \frac{\Gamma(c)\Gamma(k+1-c)}{\Gamma(k+1)} + O(t^{k-1})$$

$$= \frac{-\Gamma(k+1-c)}{c\Gamma(-c)\Gamma(k+1)} \cdot t^k + O(t^{k-1})$$

where $B(\cdot, \cdot)$ is the beta function, and we used the $\Gamma(\cdot)$ and $B(\cdot, \cdot)$ identities from pages 3 and 9, respectively, of Erdeyi (1953), volume I.

Using the above expression we get:

$$\langle T \rangle = \frac{-\Gamma(2-c)}{c\Gamma(-c)\Gamma(2)} \cdot t + O(1)$$

$$= (1-c) \cdot t + O(1)$$

$$\langle T^2 \rangle = \frac{-\Gamma(3-c)}{c\Gamma(-c)\Gamma(3)} \cdot t^2 + O(t)$$

$$= \frac{1}{2}(2-c)(1-c) \cdot t^2 + O(1)$$

so,

$$\sigma_T = \sqrt{\langle T^2 \rangle - \langle T \rangle^2}$$

$$= \sqrt{\frac{1}{2}(2-c)(1-c) \cdot t^2 - (1-c)^2 t^2 + O(1)}$$

$$= \sqrt{1-c} \cdot t \sqrt{\frac{1}{2}(2-c)+c-1+O(t^{-2})}$$

$$= \sqrt{\frac{c(1-c)}{2}} \cdot t + O(1)$$

and,

$$\frac{\sigma_T}{\langle T \rangle} = \sqrt{\frac{c}{2(1-c)}} + O(t^{-1})$$

For the case of integer valued $c$, again we observed by numerical simulations that it can be well approximated by the limiting case of the non-integer case.

In **Figures S1 and S2** we show a comparison between numerical simulations and the asymptotic exact results we derived in this section, for different values of $c$: $c = 0.5, 1.5, 2.5, 3.5$ (from top to bottom). The close agreement again demonstrates the validity of the asymptotic results in Eqs 37–40, for a wide range of timescales.

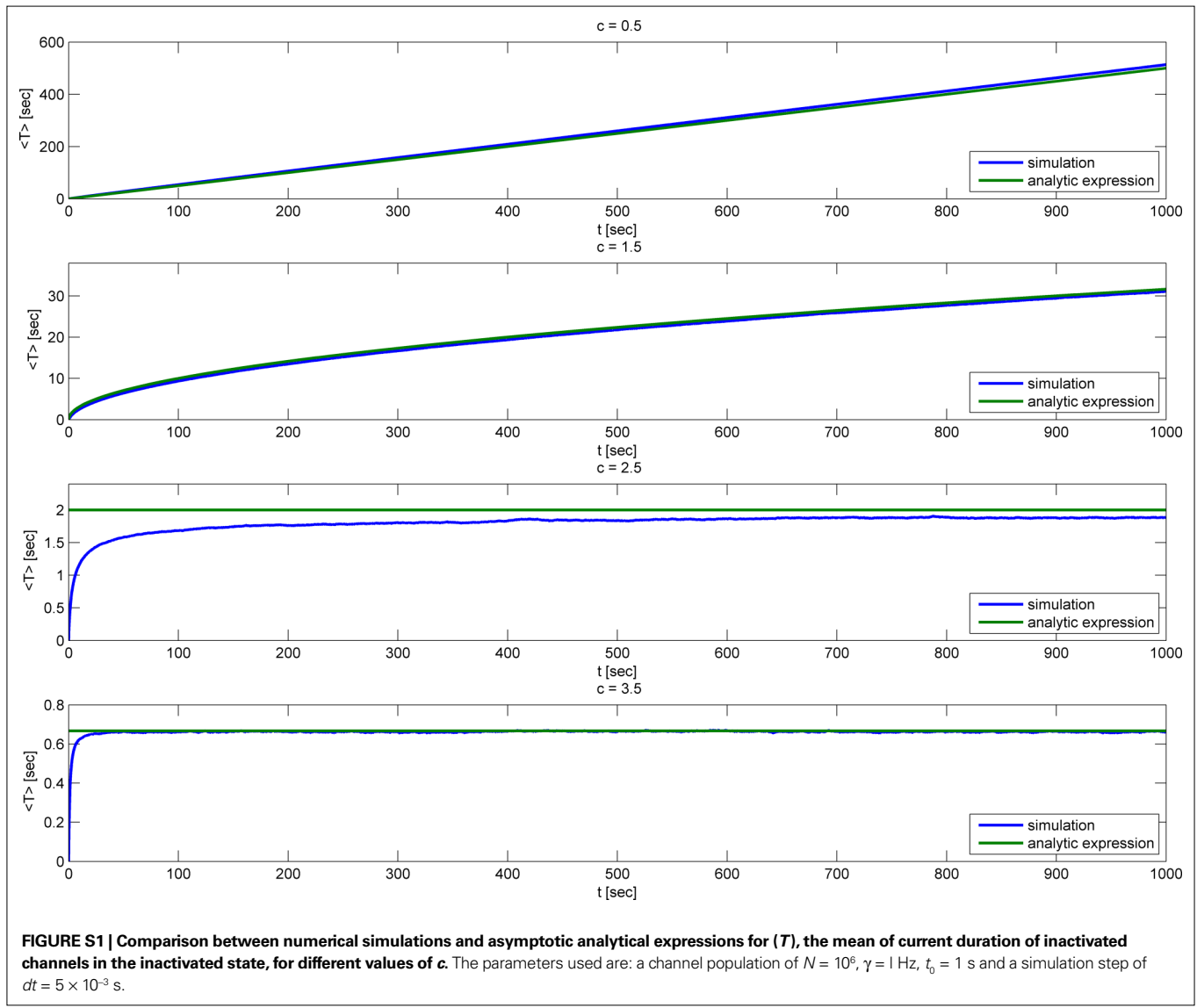

**FIGURE S1 | Comparison between numerical simulations and asymptotic analytical expressions for ⟨T⟩, the mean of current duration of inactivated channels in the inactivated state, for different values of c.** The parameters used are: a channel population of $N = 10^6$, $\gamma = 1$ Hz, $t_0 = 1$ s and a simulation step of $dt = 5 \times 10^{-3}$ s.

## WHY IS THE AVAILABLE CLUSTER OF STATE MARKOVIAN?

In Section 'Background' we mentioned that transitions between the states that compose the available state are much faster than the transitions to and from the non-available states, and therefore the available cluster of states is, to a good approximation, Markovian. In this section we use the methods presented in Section 'How to Create a Two-State Non-Markov Model from a General Markov Model' to show why this reasoning is correct.

Recall Eqs 28 and 29:

$$\psi_A(t) = \bar{\pi}_A(0) \exp\left(\mathbf{Q}_{AA}t\right)\mathbf{Q}_{AI}\bar{u}_I^T, \quad t \geq 0 \tag{S4}$$

$$\psi_I(t) = \bar{\pi}_I(0) \exp\left(\mathbf{Q}_{II}t\right)\mathbf{Q}_{IA}\bar{u}_A^T, \quad t \geq 0 \tag{S5}$$

for the RTPDFs of the 'lumped' states. When the above conditions hold, we will show that $\psi_A(t) = \gamma \exp(-\gamma t)$ for some non-negative $\gamma$, and find an explicit form for $\psi_I(t)$.

We denote by $M$ the number of available states, and by $L$ the number of inactivated states. Recall that the diagonal element of any Markovian rate matrix equals the negative sum of all non-diagonal elements. Since the transitions within the available states are much faster than transitions outside of this state, we can write:

$$\mathbf{Q}_{AA} = \tilde{\mathbf{Q}}_{AA} + \varepsilon \mathbf{D}$$

where $\tilde{\mathbf{Q}}_{AA}$ is the rate matrix for the transitions inside the available states, with all the transitions to inactivation removed from the diagonal, and $\varepsilon \mathbf{D}$ is a diagonal matrix that contains all the diagonal components of inactivation removed from $\tilde{\mathbf{Q}}_{AA}$. Formally, we can write $(\varepsilon \mathbf{D})_{ii} = -\delta_{ij}\sum_{l=1}^{L}(\mathbf{Q}_{AI})_{il}$. Since the rates of transitions to inactivation are much smaller than transitions inside the available state $\varepsilon \ll 1$ – a small parameter. Using perturbation theory (Landau and Lifshitz, 1981, chapter VI), denoting by $\{E_m\}_{m=0}^{M-1}$ and $\{\bar{\phi}_m\}_{m=0}^{M-1}$ the eigenvalues and eigenvectors of $\mathbf{Q}_{AA}$ and by $\{\tilde{E}_m\}_{m=0}^{M-1}$ and $\{\tilde{\phi}_m\}_{m=0}^{M-1}$ the eigenvalues and eigenvectors of $\tilde{\mathbf{Q}}_{AA}$, we get:

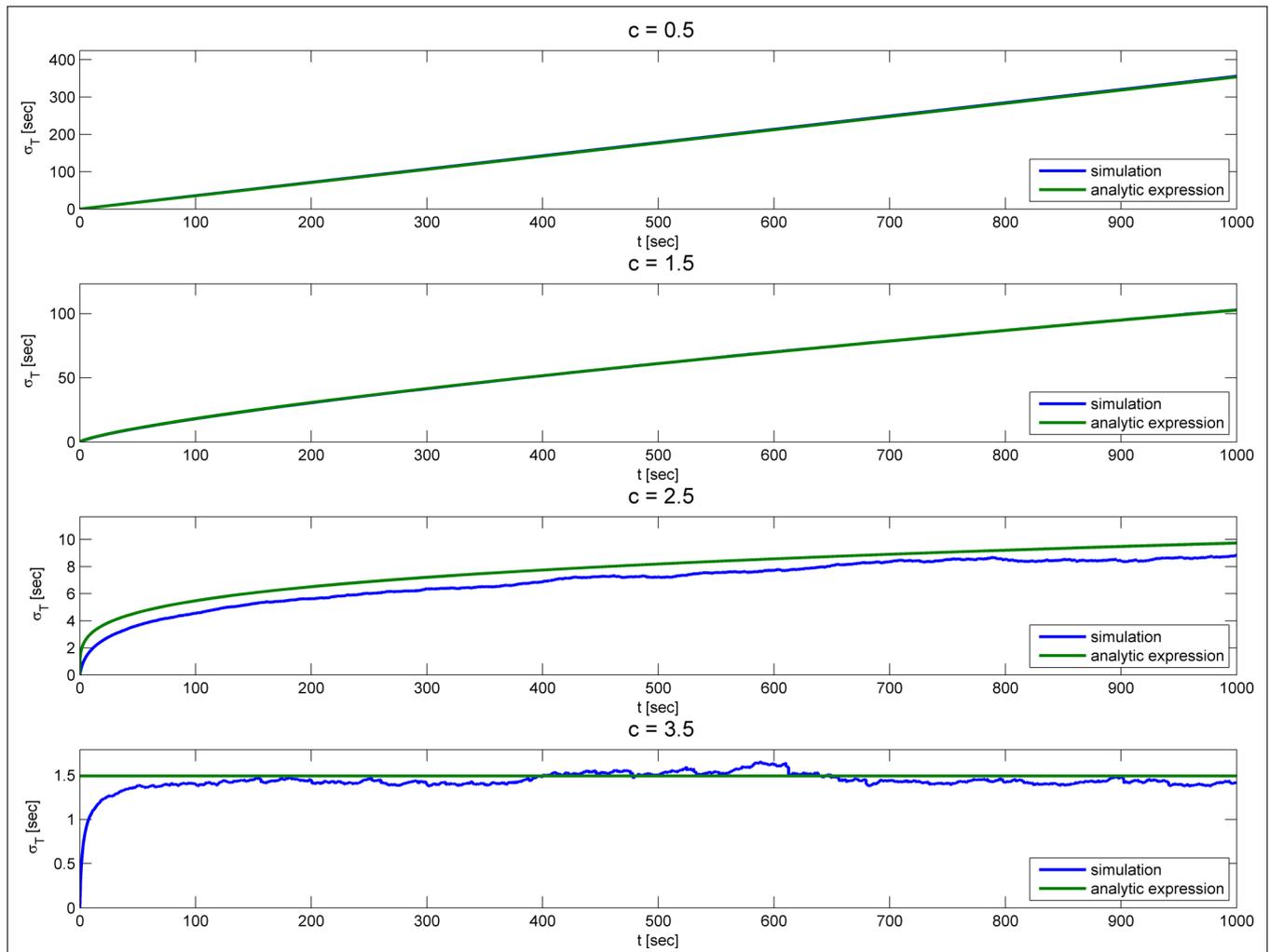

**FIGURE S2 | Comparison between and numerical simulations and asymptotic analytical expressions for $\sigma_T$, the standard deviation of current duration of inactivated channels in the inactivated state, for different values of $c$.** The parameters used are: a channel population of $N = 10^6$, $\gamma = 1$ Hz, $t_0 = 1$ s and a simulation step of $dt = 5 \times 10^{-3}$ s.

$$\bar{\phi}_m \tilde{\mathbf{Q}}_{AA} = g_m(\varepsilon)\left(\tilde{\phi}_m + \sum_{n \neq m} f_{mn}(\varepsilon)\tilde{\phi}_n\right)$$

$$E_m = \tilde{E}_m + \sum_{k=1}^{\infty} \varepsilon^k U_m^{(k)}$$

Where $f(\varepsilon) \triangleq \sum_{k=1}^{\infty} \varepsilon^k f_{mn}^{(k)}, f_{mn}^{(k)}, U_m^{(k)}$ are some constants and $g_m$ ($\varepsilon$) is a normalization constant. Since $E_m$ and $\tilde{\phi}_m$ are an eigenvalue and eigenvector of $\mathbf{Q}_{AA}$,

$$\bar{\phi}_m \mathbf{Q}_{AA} = E_m \bar{\phi}_m,$$

$$\left(\tilde{\phi}_m + \sum_{n \neq m} f_{mn}(\varepsilon)\tilde{\phi}_n\right)(\tilde{\mathbf{Q}}_{AA} + \varepsilon \mathbf{D}) = \left(\tilde{E}_m + \sum_{k=1}^{\infty} \varepsilon^k U_m^{(k)}\right)$$
$$\times \left(\tilde{\phi}_m + \sum_{n \neq m} f_{mn}(\varepsilon)\tilde{\phi}_n\right).$$

The last equation holds for all $\varepsilon$, therefore we can equate every order in $\varepsilon$ separately. For the zeroth order in $\varepsilon$,

$$\tilde{\phi}_m \tilde{\mathbf{Q}}_{AA} = \tilde{E}_m \tilde{\phi}_m \qquad (S6)$$

as expected. For the first order in $\varepsilon$,

$$\tilde{\phi}_m \mathbf{D} + \sum_{n \neq m} f_{mn}^{(1)} \tilde{\phi}_n \tilde{\mathbf{Q}}_{AA} = \tilde{\phi}_m U_m^{(1)} + \tilde{E}_m \sum_{n \neq m} f_{mn}^{(1)} \tilde{\phi}_n \qquad (S7)$$

Note that this perturbative expansion holds only in the case where the eigenvalue $\tilde{E}_m$ is non-degenerate. This applies to $\tilde{E}_0 = 0$, if we assume that $\tilde{\mathbf{Q}}_{AA}$ is indecomposable, so that it has only a single stationary distribution $\overline{\pi}_A^s$: this stationary distribution is $\tilde{\phi}_0$, the eigenvector of $\tilde{\mathbf{Q}}_{AA}$ that corresponds to the eigenvalue 0. Since $\tilde{\phi}_0$ is the only eigenvector with eigenvalue 0, and since $\tilde{\mathbf{Q}}_{AA}\overline{u}_A^T = 0$ (because $\tilde{\mathbf{Q}}_{AA}$ is a rate matrix), we get, from Eq. S6:

$$\forall m \neq 0: \tilde{\phi}_m \overline{u}_A^T = \frac{\tilde{\phi}_m \tilde{\mathbf{Q}}_{AA} \overline{u}_A^T}{\tilde{E}_m} = 0$$

while for $m = 0$, from normalization $\overline{\pi}_A^s \overline{u}_A^T = 1$.

Using this fact, we multiply Eq. S7 for $m = 0$ by $\varepsilon \overline{u}_A^T$ from the right and get:

$$\overline{\pi}_A^s(\varepsilon \mathbf{D})\overline{u}_A^T = \varepsilon U_m^{(1)}$$

This the first order correction to the 0 eigenvalue of $\overline{\pi}_A^s$ or, more simply put, the approximate eigenvalue of $\overline{\pi}_A^s$ as an approximate eigenvector of $\mathbf{Q}_{AA}$. The corrections to the other eigenvalues of $\tilde{\mathbf{Q}}_{AA}$ are negligible compared to their original, non-zero, values. Additionally, we note that we can write:

$$\begin{aligned}
\overline{\pi}_A^s(\varepsilon \mathbf{D})\overline{u}_A^T &= \sum_{i=1}^{M}\sum_{j=1}^{M}(\overline{\pi}_A^s)_i(\varepsilon \mathbf{D})_{ij} \\
&= \sum_{i=1}^{M}(\overline{\pi}_A^s)_i \sum_{j=1}^{M}(-\delta_{ij}\sum_{l=1}^{L}(\mathbf{Q}_{AI})_{il}) \\
&= -\sum_{i=1}^{M}\sum_{l=1}^{L}(\overline{\pi}_A^s)_i(\mathbf{Q}_{AI})_{il} \\
&= -\overline{\pi}_A^s \mathbf{Q}_{AI}\overline{u}_I^T
\end{aligned}$$

We set $\gamma \triangleq \overline{\pi}_A^s \mathbf{Q}_{AI}\overline{u}_I^T$ and apply these results. Using spectral decomposition, we can write:

$$\overline{\pi}_A(0)\exp(\mathbf{Q}_{AA}t) = \sum_{m=0}^{M}a_m \overline{\phi}_m \exp(E_m t)$$

where $a_m$ are constants and we recall that for every $m$, $E_m < 0$. Since for any nonzero value of $m$, $\gamma = |E_0| \ll |E_m|$, we get:

$$\sum_{m=0}^{M}a_m \overline{\phi}_m \exp(E_m t) \approx a_0 \overline{\pi}_A^s \exp(-\gamma t),$$

where $a_0 = 1$ since for $t \to 0$, $\overline{\pi}_A(0)\exp(\mathbf{Q}_{AA}t) \to \overline{\pi}_A(0)$, which must be a probability distribution (and this only occurs if $a_0 = 1$). Therefore for $t \geq 0$,

$$\begin{aligned}
\psi_A(t) &= \overline{\pi}_A^s \mathbf{Q}_{AI}\overline{u}_I^T \exp(-\overline{\pi}_A^s \mathbf{Q}_{AI}\overline{u}_I^T t), \\
&= \gamma \exp(-\gamma t).
\end{aligned}$$

Thus, we obtained an exponential distribution as in Eq. 30, and the cluster of available states is Markovian.

Next, we calculate the RTPDF of the inactivated state in this case. Assume that the channel was in the available state from time 0 to $t$ and then inactivated. From the law of total probability:

$$(\overline{\pi}_I(t))_n = \sum_{m=1}^{M}(\overline{\pi}_A(t))_m \frac{(\mathbf{Q}_{AI})_{mn}}{\sum_{l=1}^{L}(\mathbf{Q}_{AI})_{ml}}.$$

Thus, if we set $(\mathbf{G}_{AI})_{mn} \triangleq (\mathbf{Q}_{AI})_{mn}/(\sum_{l=1}^{L}(\mathbf{Q}_{AI})_{ml})$ and approximate $\overline{\pi}_A(t) = \overline{\pi}_A^s$ (zeroth order approximation according to the perturbation theory), we get:

$$\overline{\pi}_I(t) = \overline{\pi}_A^s \mathbf{G}_{AI}$$

And so, from Eq. S5,

$$\psi_I(t) = \overline{\pi}_A^s \mathbf{G}_{AI} \exp(\mathbf{Q}_{II}t)\mathbf{Q}_{IA}\overline{u}_A^T, \quad (t \geq 0)$$

As we state at Eq. 31.

## THE RENEWAL EQUATION

In this section we show how can we re-write Eq. 43 so it resembles the form of a *Renewal Equation* (see Cinlar, 1975). This expression helps gain a better understanding of the dynamic behavior.

If we multiply the general dynamic equation for the probability:

$$\frac{d}{dt}p(t) = -\gamma(t)p(t) + \int_0^t \gamma(u)p(u)\psi_P(t\,|\,u)du$$

by $\exp\left(\int_0^t \gamma(u)du\right)$, we can re-write the equation in the following way:

$$\begin{aligned}
\frac{d}{dt}\left(p(t)\exp\left(\int_0^t \gamma(z)dz\right)\right) &= \exp\left(\int_0^t \gamma(z)dz\right)\int_0^t \gamma(u)p(u)\psi_P(t\,|\,u)du, \\
p(t) &= p(0)\exp\left(-\int_0^t \gamma(z)dz\right) \\
&\quad + \int_0^t dv \int_0^v du \gamma(u)p(u)\psi_P(v\,|\,u) \\
&\quad \times \exp\left(-\int_v^t \gamma(z)dz\right).
\end{aligned}$$

Also, we define the survival probability function in the Markovian state:

$$\Psi_E(x\,|\,y) \triangleq P(t_s \geq x\,|\,t_s \geq t = u = y) = \int_x^\infty \psi_E(z\,|\,y)dz,$$

where, again, $t$ is the current time, $u$ is the time of the last transition between states, and $t_s$ is the time of the next switch in channel states.

Now we can re-write the last equation and get:

$$p(t) = p(0)\Psi(t\,|\,0) + \int_0^t dv \int_0^v du \gamma(u)p(u)\psi_P(v\,|\,u)\Psi_E(t\,|\,v) \qquad \text{(S8)}$$

This time dependent generalization of the *Renewal Equation*, has a clear intuitive meaning. This first term represents the probability that the channel has not inactivated since time $t = 0$. The second is a sum over all the probabilities for the channel to inactivate at time $u$, recover at time $v$, and remain at the available state until time $t$.

Note that in the constant input case the system becomes time invariant, and the integral with non-stationary kernels $\psi_P(v\,|\,u)$ and $\Psi_E(t\,/\,v)$ becomes a regular convolusion,

$$p(t) = p(0)\Psi_E(t) + (\gamma(t)p(t)) * \psi_P(t) * \Psi_E(t),$$

where now $\Psi_E(t) = \int_t^\infty \psi_E(z)dz$ and $*$ represent the convolution operator. A Laplace transform of this equation in the case $p(0) = 1$ leads to:

$$\tilde{p}(s) = \frac{1}{s + \gamma\left(1 - \tilde{\psi}_P(s)\right)}, \qquad \text{(S9)}$$

as expected. Note that the Laplace domain representation of the equation is limited, since we cannot generalize it to the non-stationary, varying input case, as we did in the time domain.

## THE RECOVERY RATES AND THE DYNAMIC EQUATION

In this section we show the connection between the expression for the recovery rates in Eqs 36, and 43, the dynamic equation of $p(t)$. This connection helps to understand better the meaning of $f_t(T)$.

Assume that $p(0) = 1$. The general dynamic equation for the probability, Eq. 43, can be written as the difference:

$$\frac{d}{dt} p(t) = \phi_R(t) - \phi_I(t)$$

between the flux of recovery,

$$\phi_R(t) = \int_0^t \gamma(u) p(u) \psi_P(t \mid u) du, \qquad (S10)$$

and the flux of inactivation,

$$\phi_I(t) = \gamma(t) p(t). \qquad (S11)$$

On the other hand, in this case (general input) the distribution of $T$. the time of current inactivation in the inactivated population, is (using a similar derivation to that of Eq. 36):

$$f_t(T) = (1 - p(t))^{-1} h(t - T) \Psi_P(t \mid t - T), \quad (0 \le T \le t)$$

where,

$$h(t) = \gamma(t) p(t),$$

$$\Psi_P(x \mid y) = \int_x^\infty \psi_P(z \mid y) dz.$$

Using this, and the rate of recovery,

$$\lambda_t(T) = \frac{c(t)}{T + t_0},$$

we obtain an equation for $\phi_R(t)$, by integrating over the distribution of $T$ in the inactivated population, and using Eq. 42:

$$\phi_R(t) = \int_0^t f_t(T)(1 - p(t)) \lambda_t(T) dT$$

$$= \int_0^t \gamma(t - T) p(t - T) \Psi_P(t \mid t - T) \lambda_t(T) dT$$

$$= \int_0^t \gamma(u) p(u) \psi_P(t \mid u) du$$

which is as expected, the same as Eq. S10.

## SIMULATIONS CODE

All simulations codes are available on: http://webee.technion. ac.il/~rmeir/SoudryMeir10Code.zip